\pdfoutput=1
\documentclass[11pt,twoside,a4paper,cmspaper,final,collab]{cms-tdr}
\def\svnVersion{aa052a0}\def\svnDate{2024/06/18}\def\cmsCernNoTag{CERN-EP-2024-028}\def\cmsCernDate{\today}\def\cmsMessage{Published in the Journal of High Energy Physics as \href{http://dx.doi.org/10.1007/JHEP06(2024)097}{\doi{10.1007/JHEP06(2024)097}.}}
\begin{document}\cmsNoteHeader{HIG-18-026}

\newlength\cmsFigWidth
\ifthenelse{\boolean{cms@external}}{\setlength\cmsFigWidth{0.49\textwidth}}{\setlength\cmsFigWidth{0.65\textwidth}} 
\providecommand{\cmsTable}[1]{\resizebox{\textwidth}{!}{#1}}
\newlength\cmsTabSkip\setlength{\cmsTabSkip}{1ex}

\newcommand{\higgs}{\PH}
\newcommand{\wh}{\ensuremath{\PW\higgs}\xspace}
\newcommand{\zh}{\ensuremath{\PZ\higgs}\xspace}
\newcommand{\ma}{\ensuremath{m_{\Pa}}\xspace} 
\newcommand{\mhiggs}{\ensuremath{m_{\higgs}}\xspace} 
\newcommand{\runtwolumi}{138\fbinv}
\newcommand{\htobb}{\ensuremath{\higgs\to\bbbar}\xspace} 
\newcommand{\htobbbb}{\ensuremath{\higgs\to\Pa\Pa\to\bbbar\bbbar}\xspace}
\newcommand{\pp}{\ensuremath{\Pp\Pp}\xspace}
\newcommand{\jet}{\ensuremath{\mathrm{j}}\xspace}
\newcommand{\ttjets}{\ensuremath{{\ttbar}\!+\!{\text{jets}}}\xspace}
\newcommand{\wjets}{\ensuremath{{\PW}\!+\!{\text{jets}}}\xspace}
\newcommand{\zjets}{\ensuremath{{\PZ}\!+\!{\text{jets}}}\xspace}
\newcommand{\vjets}{\ensuremath{{\PV}\!+\!{\text{jets}}}\xspace}
\newcommand{\njet}{\ensuremath{N_{\text{j}}}\xspace}
\newcommand{\nbjet}{\ensuremath{N_{\PQb}}\xspace}
\newcommand{\scalef}{\ensuremath{\mathcal{F}}\xspace}
\newcommand{\dmbb}{\ensuremath{\Delta m_{\bbbar}^{\text{min}}}\xspace}
\newcommand{\plep}{\ensuremath{p_{\text{T},\ell }}\xspace}
\newcommand{\philep}{\ensuremath{\phi_{\ptmiss,\ell}}\xspace}
\newcommand{\ttqq}{\ensuremath{\ttbar\!+\!\qqbar}\space}
\newcommand{\ttlf}{\ensuremath{{\ttbar}\!+\!{\text{LF}}}\space}
\newcommand{\ttcc}{\ensuremath{\ttbar\!+\!\ccbar}\space}
\newcommand{\ttbb}{\ensuremath{\ttbar\!+\!\bbbar}\space}
\newcommand{\tthf}{\ensuremath{{\ttbar}\!+\!{\text{HF}}}\space}
\newcommand{\btagged}{\ensuremath{\PQb\text{-tagged}}\xspace}
\newcommand{\da}{\ensuremath{D_{\mathrm{A}}}\xspace}
\newcommand{\db}{\ensuremath{D_{\mathrm{B}}}\xspace}
\newcommand{\nc}{\ensuremath{N_{\mathrm{C}}}\xspace}
\newcommand{\nd}{\ensuremath{N_{\mathrm{D}}}\xspace}
\newcommand{\iso}{\ensuremath{I_{\mathrm{iso}}}\xspace}
\newcolumntype{C}[1]{>{\centering\arraybackslash}p{#1}}

\extrafloats{200}  

\cmsNoteHeader{HIG-18-026} 
\title{Search for the decay of the Higgs boson to a pair of light pseudoscalar bosons in the final state with four bottom quarks in proton-proton collisions at \texorpdfstring{$\sqrt{s} = 13\TeV$}{sqrt(s) = 13 TeV}}

\date{\today}

\abstract{
A search is presented for the decay of the 125\GeV Higgs boson (\higgs) to a pair of new light pseudoscalar bosons (\Pa), followed by the prompt decay of each a boson to a bottom quark-antiquark pair, \htobbbb. The analysis is performed using a data sample of proton-proton collisions collected with the CMS detector at a center-of-mass energy of 13\TeV, corresponding to an integrated luminosity of~\runtwolumi. To reduce the background from standard model processes, the search requires the Higgs boson to be produced in association with a leptonically decaying \PW or \PZ boson. The analysis probes the production of new light bosons in a $15<\ma<60\GeV$ mass range. Assuming the standard model predictions for the Higgs boson production cross sections for $\pp\to\wh$ and \zh, model independent upper limits at 95\% confidence level are derived for the branching fraction $\mathcal{B}(\htobbbb)$. The combined \wh and \zh observed upper limit on the branching fraction ranges from 1.10 for $\ma = 20\GeV$ to 0.36 for $\ma = 60\GeV$, complementing other measurements in the $\PGm\PGm\PGt\PGt$, $\PGt\PGt\PGt\PGt$ and $\Pb\Pb\ell\ell$ ($\ell=\PGm,\PGt$) channels.
}

\hypersetup{%
pdfauthor={CMS Collaboration},%
pdftitle={Search for the decay of the Higgs boson to a pair of light pseudoscalar bosons in the final state with four bottom quarks in proton-proton collisions at sqrt(s)=13 TeV},%
pdfsubject={CMS},%
pdfkeywords={CMS, hadron-hadron scattering, Higgs physics, beyond standard model}}

\maketitle

\section{Introduction} 
\label{sec:introduction} 

A Higgs boson with a mass of 125\GeV was
discovered~\cite{Aad:2012tfa,Chatrchyan:2012xdj,Chatrchyan:2013lba}
in 2012 by the ATLAS and CMS Collaborations at the CERN LHC.
Studies~\cite{ATLAS:2022vkf,CMS:2022dwd} of this particle,
here denoted \higgs,
have indicated its compatibility with the Higgs boson of
the standard model (SM) of particle physics.
Notwithstanding this compatibility,
present experimental limits 
allow a significant branching fraction of the \higgs boson to beyond-the-SM 
particles up to 16\%~\cite{CMS:2022dwd}.
Observation of such \higgs boson decays could point the way
for the SM to be extended to a more fundamental theory.

Many models of physics beyond the SM predict an extended Higgs sector.
Examples include two-Higgs-doublet models (2HDM)~\cite{Branco:2011iw},
the 2HDM with a scalar singlet field (2HDM+S)~\cite{Curtin:2013fra},
and the next-to-minimal supersymmetric SM (NMSSM)~\cite{Ellwanger:2009dp}.
In the 2HDM+S model,
the Higgs sector contains seven physical scalar and pseudoscalar particles.
One of the scalar states can be identified with the 125\GeV \higgs boson
while one of the pseudoscalar states, denoted \Pa,
can be light enough to allow $\higgs\to\Pa\Pa$ decays.
These decays can occur with a significant branching fraction ($\gtrsim$10\%)
if they are kinematically allowed~\cite{Curtin:2013fra}.
For virtually all 2HDM+S scenarios,
the dominant pseudoscalar decay mode is $\Pa\to\bbbar$
(with \PQb a bottom quark)
so long as the pseudoscalar mass \ma is at least twice the bottom quark mass.

In this paper,
we present a search for the Higgs boson decay \htobbbb.
The decays are assumed to be prompt,
i.e., not leading to long-lived new particles.
The analysis is based on a sample of proton-proton (\pp) collision events at
a center-of-mass energy of 13\TeV recorded with the CMS detector in 2016--2018,
corresponding to an integrated luminosity of \runtwolumi.
The search makes use of $\PV=\PW$ or \PZ boson associated production,
in which the \PV boson decays leptonically,
$\PW\to\ell\nu$ or $\PZ\to\ell^+\ell^-$,
with $\ell$ an electron or muon.
The leptonic signature suppresses SM background processes
such as those with four \PQb jets produced
through quantum chromodynamics (QCD) interactions (multijet background).

The $\higgs\to\Pa\Pa\to\bbbar\bbbar$
channel was previously explored by the
ATLAS Collaboration~\cite{Aaboud:2016oyb,ATLAS:2018pvw,ATLAS:2020ahi}.
The present work represents the first search in this channel from the CMS Collaboration.
The ATLAS results~\cite{ATLAS:2018pvw,ATLAS:2020ahi}
were based on a data sample of about 36\fbinv at $\sqrt{s}=13\TeV$.
Thus, the present study utilizes a data sample that is approximately four times larger.
Beyond $\higgs\to\Pa\Pa\to\bbbar\bbbar$,
searches for prompt $\higgs\to\Pa\Pa$ decays have been performed in many other channels.
At $\sqrt{s}=7$ or 8\TeV,
these include 
$\higgs\to\Pa\Pa\to \Pe\Pe\Pe\Pe$~\cite{Aad:2015sva},
$\PGm\PGm\PGm\PGm$~\cite{Aad:2015sva,Chatrchyan:2012cg,Khachatryan:2015wka},
$\Pe\Pe\PGm\PGm$~\cite{Aad:2015sva},
$\PGm\PGm\PGt\PGt$~\cite{Khachatryan:2017mnf},
$\PGt\PGt\PGt\PGt$~\cite{Khachatryan:2015nba,Khachatryan:2017mnf},
$\PGg\PGg\PGg\PGg$~\cite{Aad:2015bua},
and $\Pb\Pb\PGm\PGm$~\cite{Khachatryan:2017mnf}.
At $\sqrt{s}=13\TeV$,
these include
$\higgs\to\Pa\Pa\to\Pe\Pe\Pe\Pe$~\cite{Aaboud:2018fvk},
$\PGm\PGm\PGm\PGm$~\cite{Aaboud:2018fvk,Sirunyan:2018mgs},
$\Pe\Pe\PGm\PGm$~\cite{Aaboud:2018fvk},
$\PGm\PGm\PGt\PGt$~\cite{Aad:2015oqa,Sirunyan:2018mbx,CMS:2020ffa},
$\PGg\PGg\PGg\PGg$~\cite{CMS:2022fyt,CMS:2022xxa,ATLAS:2023ian},
$\Pb\Pb\PGm\PGm$~\cite{Aaboud:2018esj,ATLAS:2021hbr,CMS:2024uru},
$\Pb\Pb\PGt\PGt$~\cite{Sirunyan:2018pzn,CMS:2024uru},
and $\gamma\gamma\,\jet\,\jet$ (where \jet is a jet)~\cite{Aaboud:2018gmx}.
These searches for the new particle \Pa cover a mass range $0.2<\ma<62.5\GeV$.
The sensitivity of these searches as a
function of \ma has recently been summarized in Ref.~\cite{Cepeda:2021rql}. 
Searches in the $\Pb\Pb\PGt\PGt$, $\PGt\PGt\PGt\PGt$, and $\Pb\Pb\Pb\Pb$
final states reach branching fraction sensitivities in the range from $1$\% to $50$\%
for $\ma \geq 2m_{\Pb} \approx 10\GeV$.

Backgrounds to \htobbbb from SM processes arise from several sources.
The dominant background arises from \ttjets production in which one or both top quarks decay leptonically.
The next-largest backgrounds are from \zjets and \wjets production  
with a \PV boson that decays leptonically: the main background process 
in this case is from $\PV+\Pb\Pb$ production plus one or two mistagged \Pb jets.
Smaller backgrounds arise from various electroweak processes.
These include events with diboson $\PV\PV^{(\prime)}$ or triboson $\PV\PV\PV^{(\prime)}$ production,
events with a single top quark,
and rare electroweak processes such as events with $\ttbar\PV$ or $\ttbar\gamma$ production.
The SM associated production of the Higgs boson in the $\PV\higgs$ channel, with \htobb,
is also expected to comprise a background component in this search.
Background from multijet production arises when jets or other reconstructed
objects are erroneously identified as leptons
or when genuine leptons are produced through hadron decays.

Tabulated results for the study are provided
in the HEPData record for this analysis~\cite{hepdata}.

\section{Detector and trigger}
\label{sec:cms} 

A detailed description of the CMS detector,
along with a definition of the coordinate system and
pertinent kinematic variables,
is presented in Ref.~\cite{Chatrchyan:2008zzk}.
Briefly,
a cylindrical superconducting solenoid with an inner diameter of 6\unit{m}
provides a 3.8\unit{T} axial magnetic field.
Within the solenoid volume
are a silicon pixel and strip tracker,
a lead tungstate crystal electromagnetic calorimeter (ECAL),
and a brass and scintillator hadron calorimeter (HCAL).
The tracking detectors cover the pseudorapidity range $\abs{\eta}<2.5$.
The ECAL and HCAL,
each with a barrel and two endcap sections,
cover $\abs{\eta}<3.0$.
Forward calorimeters extend the coverage to $\abs{\eta}<5.0$.
Muons are detected within $\abs{\eta}<2.4$ by gas-ionization chambers
embedded in the steel flux-return yoke outside the solenoid.
The detector is nearly hermetic,
permitting accurate measurements of missing transverse momentum.

The CMS trigger system is described in Refs.~\cite{Khachatryan:2016bia,CMS:2020cmk}.
For this analysis,
signal event candidates are selected using a combination of
triggers requiring either at least one electron or muon,
or,
with lower transverse momentum (\pt) thresholds on the leptons,
at least two electrons or muons.
The trigger efficiency for signal events is measured in data and is found
to exceed 95\% for events satisfying the event selection criteria
described below.

\section{Event simulation}
\label{sec:simulation} 

{\tolerance=800
With the exception of the multijet background,
which is determined from data,
the background from SM processes is evaluated from simulation
using Monte Carlo (MC) event generators.
The \ttjets background is simulated at next-to-leading order (NLO) accuracy
with the {\POWHEG}2.0~\cite{Nason:2004rx,Frixione:2007vw,Alioli:2010xd} generator.
A reweighting is applied to the top quark \pt
to improve the modeling in the \ttjets simulation.
The reweighting factors are defined as the ratio of the top quark \pt measured in data,
unfolded to the parton level,
with respect to the theory predictions at NNLO,
and are taken from Ref.~\cite{CMS:2016oae}.
The \vjets background is simulated at leading order (LO) accuracy with the
\MGvATNLO generator
(version 2.2.2 for the simulation of the 2016 data
and version 2.4.2 for the 2017--2018 data)~\cite{Alwall:2014hca},
with cross sections normalized to NLO.
For the \vjets processes,
we apply NLO electroweak corrections as a function of the \PV boson \pt
at the generator level,
as documented in Ref.~\cite{CMS:2018nsn}.
Diboson (triboson) processes are simulated at LO (NLO) accuracy
with the {\PYTHIA}8.212 ({\POWHEG}2.0) generator.
Single top quark production and the rare electroweak processes
are simulated at NLO with a combination of the {\POWHEG}2.0 and \MGvATNLO generators.
Samples generated at NLO with the \MGvATNLO generator
employ the FxFx matching and merging scheme~\cite{Alwall:2007fs},
while those generated at LO with the same generator
use the MLM matching and merging scheme~\cite{Frederix:2012ps}.
\par}

Signal events,
namely events with associated $\PV\higgs$ production followed by the decay $\higgs\to\Pa\Pa$,
are simulated at LO using the \MGvATNLO (version 2.6.5) generator.
The cross sections are normalized to the values presented in
Refs.~\cite{LHCHiggsCrossSectionWorkingGroup:2011ti,Dittmaier:2012vm,Heinemeyer:2013tqa,deFlorian:2016spz}
and include NNLO QCD and NLO electroweak corrections.
The pseudoscalar bosons are required to decay via $\Pa\to\bbbar$,
while the \PW and \PZ bosons are required to decay
through the \Pe, \PGm, or \PGt leptonic channels.
The leptonic decays of the \PGt lepton are treated as a signal contribution.
We require at least one lepton (\Pe or \PGm) with \pt greater than 10\GeV
and $\abs{\eta}$ less than 2.5: 
this results in a filter efficiency of about 61\% for the \wh channel
and 75\% for the \zh channel.
Signal event samples are generated for pseudoscalar mass values 
$\ma=15$, 20, 25, 30, 40, 50, and 60\GeV.

The parton distribution function (PDF) set used
in the simulations is NLO NNPDF3.1 for the NLO samples
and LO NNPDF3.1 for the LO samples~\cite{NNPDF:2017mvq}. 
Parton showering and hadronization are described for all MC samples
using the {\PYTHIA}8 generator.
For the simulation of the 2016 data,
the CUETP8M1 tune~\cite{Skands:2014pea,Khachatryan:2015pea} is used.
For the 2017--2018 data, we use the CP5 tune~\cite{CMS:2019csb}.
The detector response is evaluated using \GEANTfour~\cite{Agostinelli:2002hh}.
The contribution of additional \pp interactions
within the same or nearby bunch crossing (pileup)
is modeled by adding simulated minimum bias events to the MC samples,
with the distribution of additional interactions per event adjusted to agree with observation.

\section{Event reconstruction}
\label{sec:reconstruction}

The particle-flow (PF) algorithm~\cite{Sirunyan:2017ulk}
is used to reconstruct and identify individual photons, electrons, muons, charged hadrons,
and neutral hadrons (PF candidates) by combining information from individual subdetectors in an optimal fashion.
To improve the quality of the reconstruction,
requirements beyond the PF criteria are imposed on electron and muon candidates.
For electron candidates,
these requirements are based on the shower shape of the associated energy cluster in the ECAL,
the angular matching of the electron track candidate with the cluster,
and the consistency of the track with originating from the
primary \pp interaction vertex (PV)~\cite{Khachatryan:2015hwa}.
Electrons compatible with photon conversion in the detector material are removed.      
For muon candidates,
these requirements  are based on the number of energy deposits
recorded for the candidate in the silicon tracker and muon system,
the fit quality of the candidate,
and the consistency of the candidate with
originating from the PV~\cite{Sirunyan:2018fpa}.
Electron candidates are restricted to $\abs{\eta}<2.5$
and muon candidates to $\abs{\eta}<2.4$.

The PV is taken to be the vertex corresponding
to the hardest scattering in the event,
evaluated using tracking information alone as described in Section 9.4.1 of Ref.~\cite{CMS-TDR-15-02}.
The PV is required to lie within 24\cm of the center of the detector
in the direction along the beam axis and within 2\cm in the plane
transverse to that axis.

Leptons from \PV boson decays are generally isolated from the jets in an event,
unlike leptons from sources such as hadron decays.
We therefore require electron and muon candidates to be isolated.
An isolation variable \iso is defined as the sum of the scalar \pt of charged hadron,
neutral hadron, and photon PF candidates within a cone of radius
$\Delta R=\sqrt{\smash[b]{(\Delta\eta)^2+(\Delta\phi)^2}} = 0.3$
around the electron or muon candidate direction,
where $\phi$ is the azimuthal angle. 
A correction is applied to account for the contribution of pileup.
For electrons,
this correction is performed using the jet area method
described in Ref.~\cite{Cacciari:2007fd}.
For muons,
the correction is performed by subtracting half the sum of scalar \pt
of charged-particle tracks within the cone,
but using only those tracks that do not originate from the PV.
We require the ratio of \iso to the \pt value of the muon 
to be less than 0.15. 
For electrons, the \iso selection requirements are optimized via a multivariate analysis,
and are chosen to be 0.059 ($0.029 + 0.506\pt$)
and 0.057 ($0.044+0.963\pt$) in the ECAL barrel and endcap, respectively,
for the 2016 (2017--2018) data-taking period.

The combined reconstruction, identification, and isolation efficiency
for electrons and muons is determined in data using \PZ boson decays
with a tag-and-probe method~\cite{CMS:2011aa}.
The efficiencies range from 70 to 90\% for electrons
and from 90 to 97\% for muons,
depending on the lepton \pt and $\eta$. 
 
Jets are reconstructed by clustering PF candidates using
the anti-\kt jet algorithm~\cite{Cacciari:2008gp,Cacciari:2011ma}
with a distance parameter of~0.4.
Charged-particle tracks not associated with the PV
are excluded from the clustering. 
The jet energies are corrected as explained in Ref.~\cite{Khachatryan:2016kdb}
to account for the nonlinear response of the detector and
for the expected contributions of neutral particles from pileup.
The jets are required to satisfy $\pt>20\GeV$ and $\abs{\eta}<2.5$. 
The number of selected jets in an event is denoted \njet.

Bottom quark jets are identified by applying
the \textsc{DeepCSV} algorithm~\cite{Sirunyan:2017ezt} to the selected jets
(the \pt threshold is maintained at 20\GeV).
The medium working point of this algorithm is used, 
for which the tagging efficiency is 68\% for \PQb quark jets with $\pt\approx30\GeV$.
The corresponding misidentification probability for light jets
(\ie, gluon and light quark jets) is 1.6\%,
while that for charm quark jets is 13\%.
The number of \btagged jets in an event is denoted~\nbjet.

The missing transverse momentum vector \ptvecmiss is defined as the projection 
onto the plane perpendicular to the beam axis of the negative vector sum of the 
momenta of all reconstructed PF candidates in an event.
Its magnitude is referred to as \ptmiss.
Calibration and pileup effects are accounted for
as described in Ref.~\cite{Sirunyan:2019kia}.
The \ptmiss variable is used to measure the undetected
momentum in an event,
such as is carried by neutrinos in leptonic \PW boson decays
or as can arise spuriously from the misreconstruction of jet \pt.

For the \wh channel
we also consider the transverse mass \mT, defined by
\begin{equation}
  \mT=\sqrt{2\ptmiss\plep(1-\cos\philep)},
\end{equation}
where \plep is the transverse momentum of the lepton
and \philep is the azimuthal angle between \ptvecmiss and the \plep vector.
For \wh events,
\mT represents the \PW boson mass calculated
using transverse momentum only~\cite{Arnison:1983rp}.

\section{Analysis procedure}
\label{sec:procedure}

\subsection{Baseline selection}
\label{sec:selection}

A set of baseline selection criteria are applied in order to identify events
that are consistent with the signal processes considered.

Single-lepton triggers are used to identify events in the \wh channel.
The \pt threshold for electrons was 27\GeV in 2016 and 32\GeV in 2017--2018.
The corresponding thresholds for muons were 24 and 27\GeV, respectively.
Dilepton triggers are used to identify events in the \zh channel.
For electrons, the \pt thresholds on the two leptons were 23 and 12\GeV,
while for muons they were 17 and 8\GeV,
for all data-taking periods.
For muons,
the triggers in 2017--2018 differ from those in 2016
by the additional requirement that the dimuon invariant mass be larger than 3.8\GeV.
The triggers require the leptons to satisfy stringent identification criteria
and to be isolated from other tracks
and energy deposits in the calorimeters.

For the \wh channel,
events must contain exactly one electron or muon, and 
no other electron (muon) with \pt above 15 (10)\GeV.
For the 2016 (2017--2018) data,
electrons must have $\pt>30$ (35)\GeV and muons $\pt>25$ (30)\GeV. 
Higher thresholds are used for the 2017--2018 data to account for the 
increase in the online trigger thresholds.
To suppress the multijet background,
\wh events must satisfy $\ptmiss>25\GeV$ and $\mT>50\GeV$.

For the \zh channel,
events must contain exactly one \EE or \MM pair 
and no other electron or muon with \pt above 15 (10)\GeV.
For electron pairs,
one of the two legs must have $\pt>25\GeV$ and the other $\pt>15\GeV$.
For muon pairs,
the corresponding requirements are $\pt>20\GeV$ and $\pt>10\GeV$.
To reduce the background from \ttjets events,
the invariant mass of the lepton pair must lie within
20\GeV of the nominal \PZ boson mass.
No requirements are imposed on \ptmiss or \mT.

For both \wh and \zh events, we require $\njet\ge3$ and $\nbjet\geq2$.

\subsection{Analysis overview}
\label{sec:strategy} 

The selected events are categorized in terms of \nbjet and \njet.
Events in the signal region (SR) are divided into two categories depending on whether
they have $\nbjet=3$ (denoted the 3b category) or~$\nbjet=4$ (denoted the 4b category).
The 4b category corresponds to events in which all four \Pb jets in
\htobbbb decays
have been reconstructed.
The 3b category accounts for events in which two \Pb quarks are merged in a single 
jet or one of the \Pb quarks is rather soft and fails to be reconstructed.
The latter case is expected to have a small impact on the Higgs boson kinematics.

For \wh events, the signal region is defined by
(\nbjet, $\njet) = (3$b, 3--4j) and (4b, 4j),
where 3--4j refers to events with either exactly three or exactly four jets.
For \zh events,
the SR is defined by (3b, $\geq$3j) and (4b, $\geq$4j).
These regions are summarized in Table~\ref{tab:crs}.
The \higgs boson is formed from the four \btagged jets
in the 4b category and from the three \btagged jets in the 3b category.
For \wh events,
the \PW boson is defined by the electron or muon
and the \ptvecmiss vector.
For \zh events,
the \PZ boson is defined by the $\EE$ or $\MM$ pair.

\begin{table}[tb] 
\centering  
\topcaption{
Signal region (SR) and control region (CR) requirements in (\nbjet, \njet) for the \wh and \zh channels,
where \nbjet is the number of \btagged jets in an event and \njet is the total number of jets in an event.
}
\label{tab:crs}
\begin{tabular}{ccc}
  \hline
Label   & (\nbjet, \njet)  & Description \\
  \hline
  \multicolumn{3}{c}{\wh channel} \\
  SR (3b)            &  (3b, 3--4j)      &  3b signal region \\
  SR (4b)            &  (4b, 4j)         &  4b signal region  \\
  CR (3b)            &  (2b, 3j)         & \PW/\ttjets control region \\
  CR (4b)            &  (2b, 4j)         & \ttjets control region \\[\cmsTabSkip]
  \multicolumn{3}{c}{\zh channel} \\
  SR (3b)            &  (3b, $\geq$3j)   &  3b signal region \\
  SR (4b)            &  (4b, $\geq$4j)   &  4b signal region \\
  CR (3b)            &  (2b, 3j)         &  DY control region \\
  CR (4b)            &  (2b, 4j)         & DY control region \\
  \hline
\end{tabular}
\end{table}           

A multivariate discriminator based on a
boosted decision tree (BDT)~\cite{Hocker:2007ht}
is used to help separate signal and background events. 
Different BDTs are constructed for the \wh and \zh channels, and the 3b and 4b event categories.
The input variables for the BDT are:
\begin{itemize}
\item $\pt^{\higgs}$:
  the vector sum of the transverse momenta of the three or four \btagged jets forming the \higgs boson candidate;
\item \mhiggs:
  the invariant mass of the three or four \btagged jets forming the \higgs boson candidate;
\item $\pt^\PV$:
  for \PW boson candidates, the magnitude of the vector sum of the \pt of the electron or muon and \ptmiss;
  for \PZ boson candidates, the \pt of the dilepton pair;
\item $\HT$:
  the scalar sum of the \pt of the three or four \btagged jets that define the \higgs boson candidate;
\item $\langle\Delta R(\PQb,\PQb')\rangle$:
  the separation in the $\eta$-$\phi$ plane between any two
  \btagged jets in an event,
  averaged over all such combinations;
\item $\abs{\Delta \phi(\PV,\higgs)}$:
  the azimuthal angle between the directions of the \PW or \PZ
  and the \higgs boson candidates;
\item $\abs{\Delta \phi(j,\ptmiss)}^{\text{min}}$:
  the smallest azimuthal angle between the \ptvecmiss direction and a jet;
\item $\pt^\ell$: 
  for \wh events, the \pt of the electron or muon;
  for \zh events, the \pt of the leading electron or muon;
\item \ptmiss: the magnitude of the missing transverse momentum;
\item \mT (defined only for \wh events): the transverse mass defined in Section~\ref{sec:reconstruction};
\item $\dmbb=\abs{m_{\bbbar,1}-m_{\bbbar,2}}^{\text{min}}$
  (defined only for the 4b category):
  the minimum difference between two dijet masses formed
  from all possible combinations with the four \btagged jets.
  The four \btagged jets can be grouped in three different ways to form a pair of $\Pa\to\bbbar$ candidates.
  The \dmbb variable represents the grouping that results in the smallest difference
  between the masses of the two pseudoscalar candidates.
\end{itemize}

The BDTs are trained on signal events, combining all pseudoscalar boson mass points,
and on background events, combining all considered SM processes.
The shapes of the BDT distributions 
are not strongly dependent on the assumed value of \ma,
allowing the same BDT to be used for all signal mass values.

We perform a maximum likelihood fit of the signal and background
components to the observed distribution of the BDT discriminant.
The fit is performed simultaneously across the SRs and control regions (CRs).
The CRs are described in Section~\ref{sec:background}.
The templates for signal events are derived from simulation.
The templates for background events are determined as
described in Section~\ref{sec:background}.

\subsection{Background modeling and control regions}
\label{sec:background}

A major source of background in our study is from \ttjets production,
as mentioned in the introduction.
To evaluate this background,
simulated \ttjets events are categorized based on
the parton-level flavor content of additional particle-level jets not originating
from the decay of the \ttjets system.
Events that contain at least one additional particle-level jet matched 
to a \PQb hadron are labeled as \ttbb.
Events that contain at least one additional particle jet matched to a \PQc hadron
and no \PQb hadron are labeled as \ttcc.
The \ttcc and \ttbb categories, with heavy-flavor (HF) production, 
are referred to as the \mbox{``{\tthf}''} background.
This background mostly arises from events with gluon splitting, $\Pg\to\ccbar$ or $\bbbar$.
The remaining \ttjets events are referred to as the \mbox{``{\ttlf}''} background, 
where ``LF'' refers to light flavor.
The \ttlf category also includes events 
with no additional particle-level jets. One source of this background is 
\ttqq ($\PQq = \PQu$, \PQd, \PQs) events in which light-flavored jets are 
erroneously identified as \PQb jets.

The template for the \ttlf background is taken from simulation,
both for its shape and normalization. In contrast to \ttlf events, 
the rate of \tthf events is not well predicted by simulation because 
it is dominated by the gluon splitting process,
which is not yet well understood to adequate precision~\cite{ATLAS:2016anw}.
Therefore, while the shapes of the templates and the relative rate of \ttbb to \ttcc 
production are taken from simulation, the overall normalization is 
determined from the fit to data. 

The \wjets and \zjets backgrounds mostly arise
when a \PW or \PZ boson is produced in association with \PQb quark jets.
Like \tthf production,
the rates of these processes are difficult to model~\cite{Aad:2014dvb,CMS:2020hmf}.
Thus,
for the \wjets background to the \wh channel
and the \zjets background to the \zh channel,
the shapes of the templates are taken from simulation while the normalizations
are determined in the fit.
There is a small \zjets background to the \wh channel.
The template for this background is taken from simulation
for both the shape and normalization.
The \wjets background to the \zh channel is negligible and is therefore ignored.

The CRs used to constrain the normalizations of the \tthf, \wjets, and \zjets processes
are obtained by selecting events with one less \btagged jet with respect to the~SRs.
This requirement enables a collection of events predominantly 
governed by these background processes,
while simultaneously depleting the signal content but closely aligning with the~SRs.
The (\nbjet, \njet) criteria for the CRs are as listed in Table~\ref{tab:crs}.
In the (2b, 3j) and (2b, 4j) categories,
any variable that depends on \nbjet is constructed by adding ``fake'' \Pb tags
using the untagged jet(s) with highest \textsc{DeepCSV} discriminator value(s). 
For the \wh channel,
we define a \PW/\ttlf CR in the \mbox{(2b, 3j)} category and a \ttlf CR in the (2b, 4j) category.
For the \zh channel,
we define a \zjets CR in the (2b, 3j) and (2b, 4j) categories.
The modeling of the \tthf background is validated in signal-depleted regions of the SR bins
corresponding to BDT scores~$\lesssim$0.1. 
The modeling of the various backgrounds has been studied through comparison with observations in the CRs.
The degree of agreement for shape variations 
between CR data and pre-fit estimates is validated to within $\approx$10\%.

For the small backgrounds from diboson, triboson,
single-top quark, rare electroweak processes,
and the SM $\PV\higgs$ process with \htobb,
both the shape and normalization are taken from simulation. 
Because of the presence of two isolated leptons,
the multijet background to the \zh channel is negligible and is ignored.
In the \wh channel,
for which only one isolated lepton is required,
the multijet background is expected to be small.
We estimate this background contribution from data,
as described below.

To evaluate the multijet background to the \wh channel,
we select data events by inverting the lepton isolation requirements
of Section~\ref{sec:reconstruction}.
All other selection criteria are the same as for signal events.
We call this sample the inverted-isolation sample.
The multijet background template is given by the measured BDT distribution
of events in the inverted-isolation sample, normalized by a scale factor \scalef
calculated in a low-\ptmiss data sideband.
The low-\ptmiss sideband is
defined by $\ptmiss<25\GeV$ and $\mT<50\GeV$.
We choose a sideband in \ptmiss and \mT because,
for multijet background events,
these variables are largely uncorrelated with the lepton isolation variable.

We thus consider the following four quantities:
\begin{itemize}
  \item \da: The BDT distribution of multijet background events in the SR;
  \item \db: The BDT distribution of multijet background events in the inverted-isolation sample;
  \item \nc: The number of multijet background events in the low-\ptmiss sideband that
    satisfy the lepton isolation requirements;
  \item \nd: The number of multijet background events in the low-\ptmiss sideband that
    fail the lepton isolation requirements.
\end{itemize}
The multijet background estimate is given by
$\da = \scalef\,\db$ with $\scalef\equiv \nc/\nd$.
The contributions of nonmultijet background events to
\db, \nc, and \nd are subtracted using simulation before calculating~\da.
Because the multijet background is expected to be small compared to other backgrounds,
and because the statistical precision in this background estimate is limited,
the estimate is assigned a conservative uncertainty of 50\%.
The closure or self-consistency of the estimate is tested in the 3b and 4b CRs,
where the contribution of multijet events is small.
The predicted multijet background in these CRs is found to be in reasonable agreement
with the data,
to within the assigned uncertainty,
once the nonmultijet contributions in data have been subtracted.

The same procedure is used to account for the multijet background contributions
to the \tthf and \wjets CRs for the \wh channel.

\subsection{Systematic uncertainties} 
\label{sec:systs} 

The overall uncertainty in the determination of the integrated luminosity
over the 2016--2018 data-taking period
is~1.6\%~\cite{CMS:2021xjt,CMS-PAS-LUM-17-004,CMS-PAS-LUM-18-002}.
Systematic uncertainties related to the event simulation include those
associated with the PDFs,
the value of the strong coupling strength \alpS,
and the renormalization and factorization scales.
We evaluate these uncertainties by altering the PDFs and scales of the simulated background
samples according to the PDF4LHC prescription of Ref.~\cite{Butterworth:2015oua}.
Their effect on the yields is estimated to be 4--5\%.
The differences between the NNPDF3.1 and PDF4LHC uncertainties are found to be small 
compared to the statistical uncertainties in the background templates.
The uncertainties in the signal cross section arising from the perturbative QCD calculations,
PDF, and the \alpS scales have a negligible effect on the shape of the signal.
Their effects on the yield correspond to 0.7 and 1.9\%, respectively, for \wh production,
and to 3.8 and 1.6\%, 
respectively, for \zh production~\cite{LHCHiggsCrossSectionWorkingGroup:2011ti},
and are taken into account by introducing nuisance parameters with
log-normal distributions into the fit. 
We assign the following theoretical uncertainties on 
the cross sections for those simulated backgrounds
whose normalization is not determined in the fit:
\ttlf events (6\%)~\cite{Czakon:2013goa}, single top quark events (5\%)~\cite{Kidonakis:2010ux, Kidonakis:2013zqa},
diboson and triboson events (50\%)~\cite{Gehrmann:2014fva,Cascioli:2014yka}, 
\zjets events in the \wh channel (2\%)~\cite{CMS:2018mdl},
and rare electroweak processes (15\%).

Tables~\ref{tab:statsyst_wh} and~\ref{tab:statsyst_zh}
summarize the other systematic uncertainties that are considered in the analysis. 
For those, the shape variations are taken into account in the template fit, while their effect
on the background and signal yield has been evaluated for illustration purposes.
For signal events, the estimate of these uncertainties are shown for two choices 
of the pseudoscalar mass: $\ma=20$ and 60\GeV.

The uncertainties related to the trigger efficiency
and the lepton identification and isolation requirements are evaluated using data
(this latter uncertainty, at 2\%,
is the same for electrons and muons).
The uncertainties associated with the lepton energy scale, the jet energy scale,
and the jet energy resolution are evaluated as a function of lepton or jet \pt and~$\eta$.
The uncertainty associated with the lepton energy scale is found to be negligible.
In Tables~\ref{tab:statsyst_wh} and~\ref{tab:statsyst_zh},
the uncertainty quoted for the jet energy scale is the
maximum that is found for the various \pt and $\eta$ bins.
A related uncertainty,
relevant for \wh channel,
is in the energy scale of unclustered \ptmiss,
where unclustered \ptmiss refers to signals in the detector
that are not associated to reconstructed physics objects, namely muons, electrons, 
photons, hadronically decaying taus and jets~\cite{Sirunyan:2019kia}.
The uncertainties in the efficiencies for \PQb tagging
and for the mistagging of \PQc quark and light-flavored jets
are evaluated as described in Ref.~\cite{Sirunyan:2017ezt}.
Sources of systematic uncertainty that affect the data-to-simulation corrections
of the \Pb-tagging efficiency and misidentification rate
are the jet energy scale,
the cross contamination between light flavor or gluon jets with heavy flavor jets,
and the statistical fluctuations in the data and MC.
These uncertainties are split into 15 \pt and $\eta$ bins
and consist of 9 independent sources.
Except for their statistical component,
they are treated as correlated between the three data-taking periods. 
To evaluate the uncertainty associated with pileup,
the total inelastic \pp cross section is varied by~$\pm$5\%~\cite{Sirunyan:2018nqx}. 
The uncertainty from the mismodeling of the top quark \pt is assessed 
for the \ttjets MC by comparing results with and without reweighting.
This uncertainty is negligible compared to the other uncertainties in the \zh channel.

\begin{table}[htp]
\centering
\topcaption{
Summary of systematic uncertainties and their effect on the background and signal event
yields in the \wh channel.
Uncertainties that are negligible are indicated with a dash~(\NA).
}
\label{tab:statsyst_wh}
\cmsTable{
\begin{tabular}{lccccccc}\hline
\multirow{2}{*}{Source}    & \multicolumn{7}{c}{Uncertainty (\%)}  \\
                                     & \ttbb & \ttcc & \ttlf & \wjets & \zjets & \wh (20\GeV) & \wh (60\GeV) \\\hline
Lepton trigger, identification \& isolation   & 2    & 2    & 2    & 2   & 2 & 2   & 2 \\
Jet energy scale                     & 0.5  & 0.5  & 0.5  & 2   & 2 & 1 & 1 \\
Jet energy resolution                & 0.5  & 0.5  & 0.5  & 4   & 5  & 1 & 0.5 \\
Energy scale of unclustered \ptmiss  & 0.5  & 0.5  & 0.5  & 2 & 2  & 1 & 0.8 \\
\PQb quark tagging                   & 6    & 4  & 4  & 4.5 & 5  & 9 & 8.5 \\
\PQc quark mistagging                & 1.5    & 6.5  & \NA    & 4 & 2.5  & 0.3 & 0.1 \\
Light-flavor mistagging              & 1.5    & 1  & 9  & 7 & 9.5  & 1 &  0.5 \\
Limited MC statistical precision     & 6    & 1.5  & 1  & 8 & 7  & 1 & 1 \\
Pileup                               & 0.5  & 1  & 2.5  & 4   & 4 & 0.5 & 0.5 \\
\PQt quark \pt mismodeling           & 2.5  & 1.3 & 1.5 & \NA & \NA & \NA & \NA \\ \hline
\end{tabular}
}
\end{table}

\begin{table}[htp]
\centering
\topcaption{
Summary of systematic uncertainties and their effect on the background and signal event
yields in the \zh channel.
Uncertainties that are negligible are indicated with a dash~(\NA).
}
\label{tab:statsyst_zh}
\cmsTable{
\begin{tabular}{lcccccc}\hline
\multirow{2}{*}{Source}    & \multicolumn{6}{c}{Uncertainty (\%)}  \\
                                    & \ttbb & \ttcc  & \ttlf & \zjets & \zh (20\GeV) & \zh (60\GeV) \\\hline
Lepton trigger, identification \& isolation  & 2    & 2    & 2    & 2       & 2              & 2 \\
Jet energy scale                    & 1    & \NA  & \NA  & 0.5  & 0.1 & 0.1 \\
Jet energy resolution               & 0.1    & 0.1  & 0.1  & 0.1  & 0.1 & 0.1 \\
\PQb quark tagging                  & 6.5    & 4  & 4  & 5 & 9 & 8 \\
\PQc quark mistagging               & 0.5    & 9  & \NA & 2.5  & 0.3 & 0.1 \\
Light-flavor mistagging             & 1    & 2  & 10  & 6  & 1 & 0.5 \\
Limited MC statistical precision    & 8    & 5  & 1  & 5  & 1 & 1 \\
Pileup                              & 1  & 1  & 4  & 3  & 0.5 & 0.5 \\ \hline
\end{tabular}
}
\end{table}

To account for the finite sizes of the simulated event samples,
each bin of the simulated signal-plus-background templates is allowed
to vary within its statistical uncertainty, 
independently from the other bins in the distribution.
A  bin-by-bin statistical uncertainty is considered by assigning a Poisson 
nuisance parameter per bin to distributions in those samples~\cite{Barlow1993}. 

\subsection{Maximum likelihood fit}
\label{sec:fit}

A maximum likelihood fit of the signal and background terms to the observed
distribution of the BDT discriminant is performed
simultaneously across the SRs and CRs.
The fits are performed separately for the \wh and \zh channels
and also with the \wh and \zh channels combined.

For the \wh channel,
there are two SRs and two CRs,
as indicated in the upper portion of Table~\ref{tab:crs},
for a total of four independent data regions.
In addition,
the electron and muon samples are treated as separate event categories in the fit.
We also treat each of the three years of  data collection separately.
There are therefore 24 data regions that are fitted simultaneously.
The fitted parameters are the signal strength modifier,
a global scale factor for \tthf production,
a global scale factor for \wjets production,
and various nuisance parameters,
where the nuisance parameters represent the systematic uncertainties from Section~\ref{sec:systs}.
The signal strength modifier $\mu$ is defined by
the equation: $\mathrm{s} = \mu \mathrm{s_0}$, where $\mathrm{s}$ is the observed signal yield 
and $\mathrm{s_0}$ is the expected signal yield for a SM $\PV\higgs$ production cross section 
times a ``benchmark'' branching fraction $\mathcal{B}(\htobbbb) = 1$. 

For the \zh channel,
there are also two SRs and two CRs,
as indicated in the lower portion of Table~\ref{tab:crs},
leading to 24 independent data regions in the fit as for the \wh channel.
The fitted parameters are the signal strength modifier,
a global scale factor for \tthf production,
a global scale factor for \zjets production with three \btagged jets,
a global scale factor for \zjets production with four \btagged jets,
and several nuisance parameters.
We implement separate scale factors for \zjets production
with three and four \btagged jets
because the relative rates of \PZ boson production in association
with two or more \PQb quarks
are not yet well established~\cite{Aad:2014dvb,CMS:2020hmf}.
For \wjets events,
a common scale factor is used
for the three and four \btagged categories
because the large statistical uncertainties do not allow discrimination
between the two categories.

Systematic uncertainties in the signal and background normalizations
are treated as log-normal distributions. The SR and CR distributions
in data and the background estimates are
used in the maximum likelihood fit to search for \htobbbb decays. 

\section{Results} 
\label{sec:results} 

\begin{figure}[t!]   
\centering   
\includegraphics[width=0.47\textwidth]{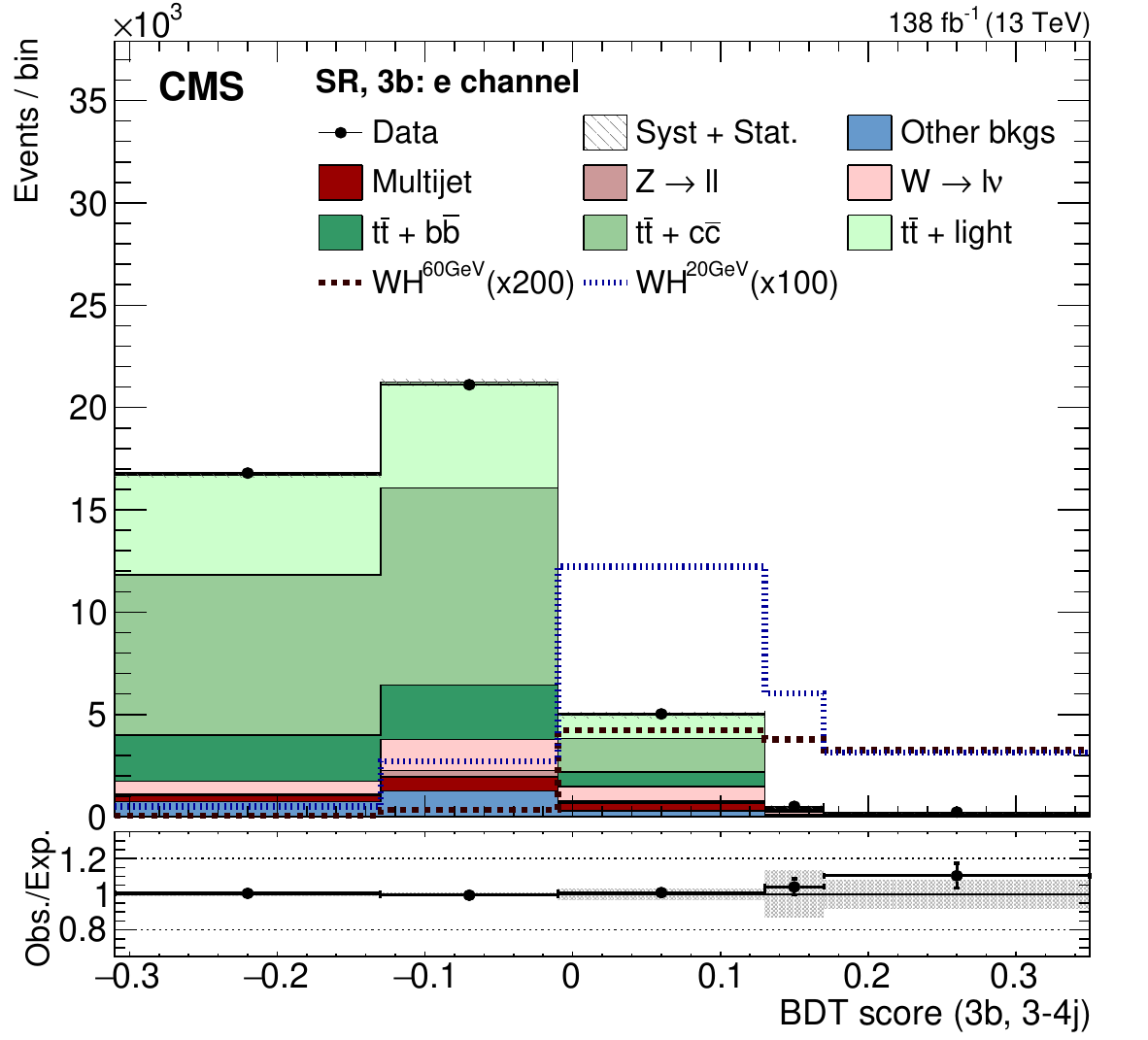}   
\hspace{3mm}
\includegraphics[width=0.47\textwidth]{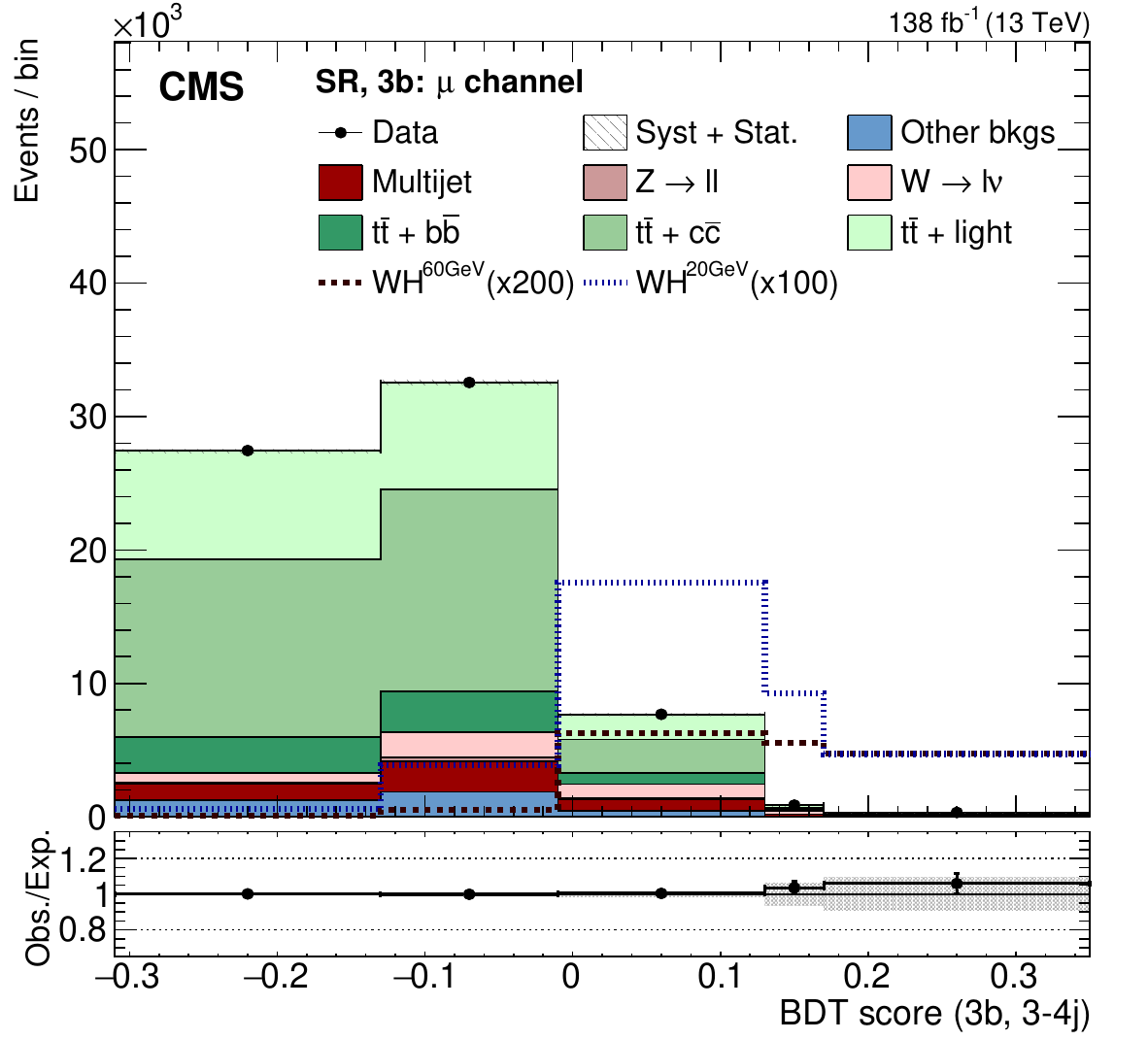} 
\includegraphics[width=0.47\textwidth]{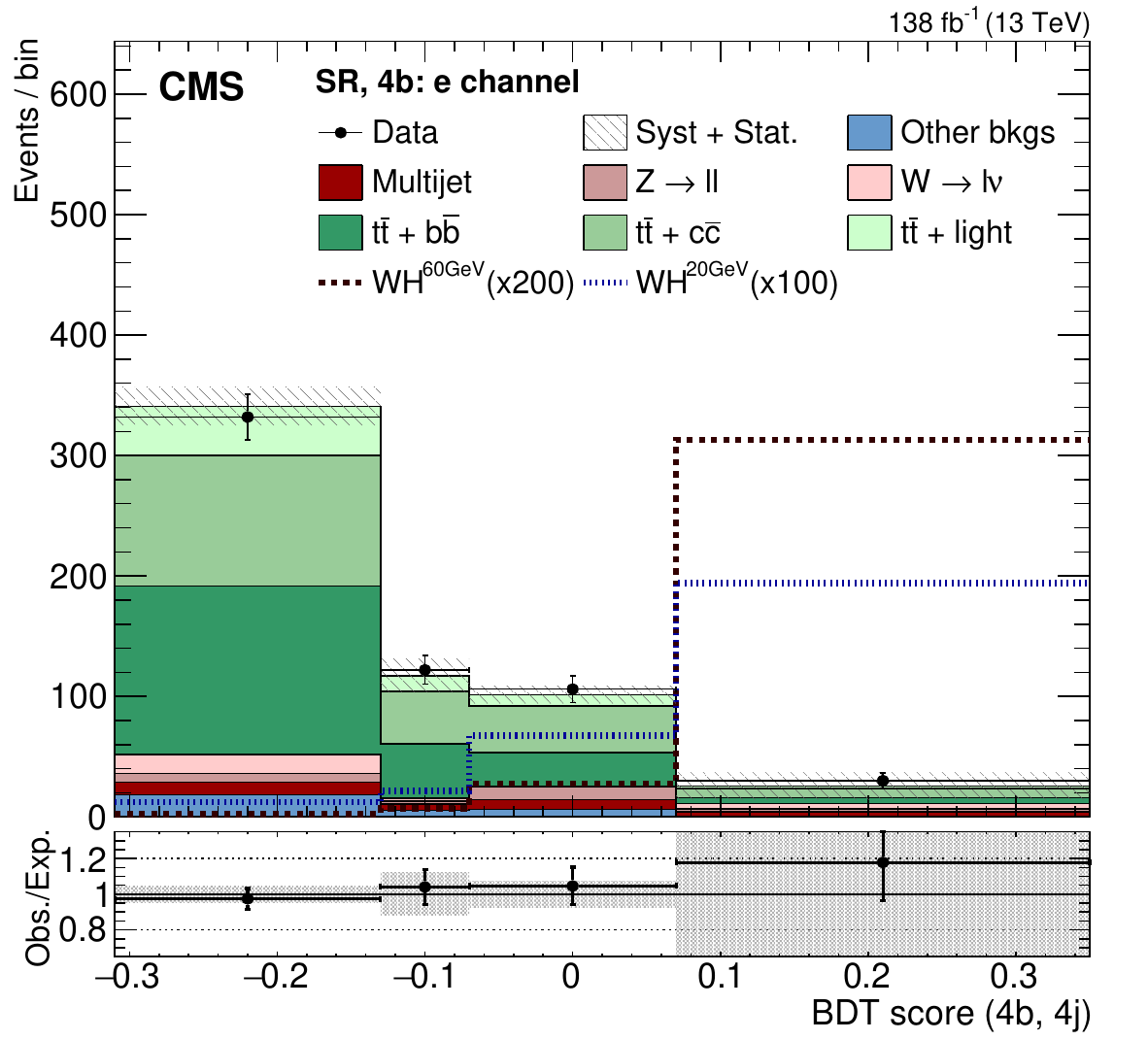}   
\hspace{3mm}  
\includegraphics[width=0.47\textwidth]{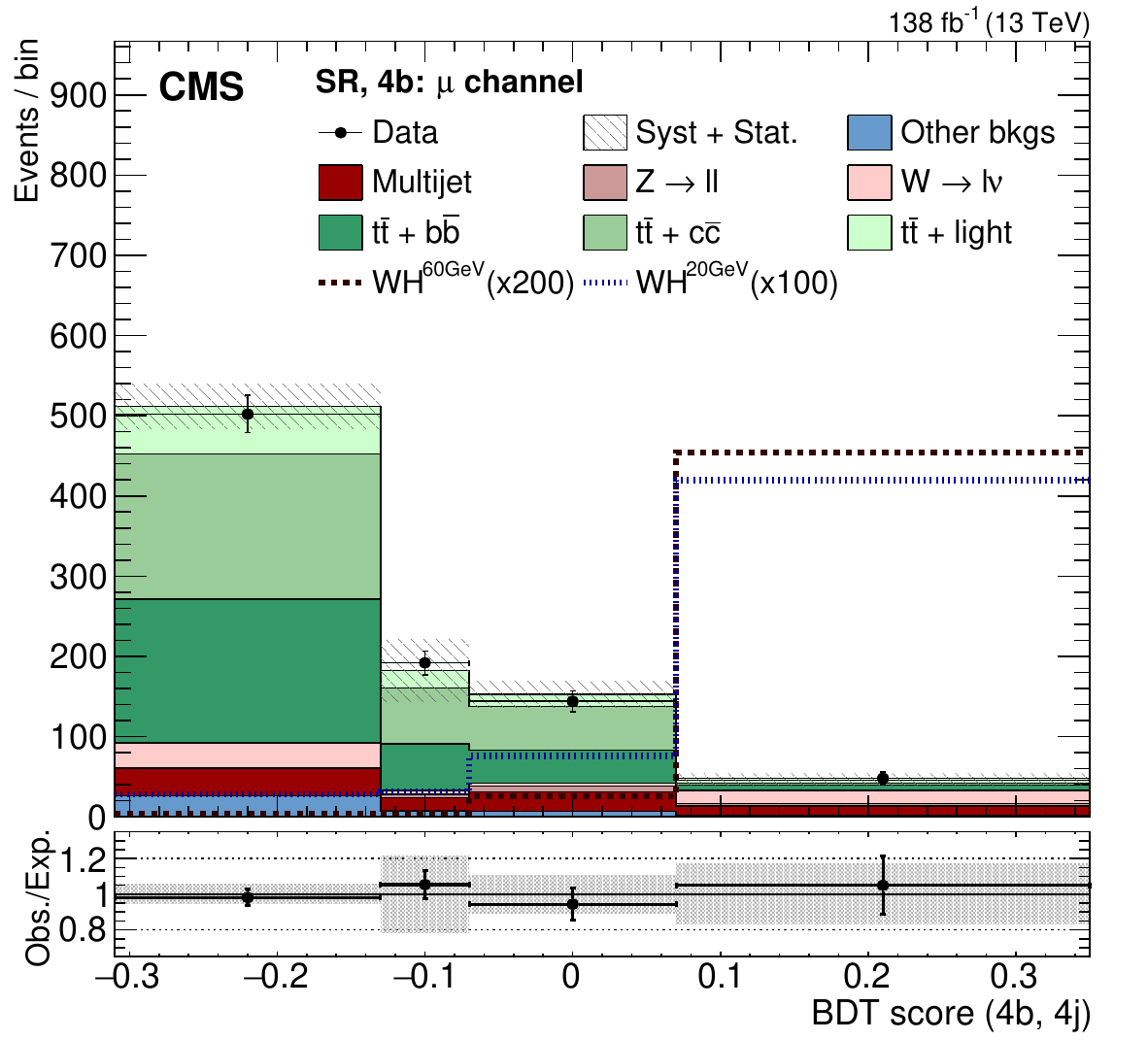}
\vspace{4mm}
\caption{
Post-fit BDT distributions in the \wh channel
extracted with the $\ma=60\GeV$ signal hypothesis.
Signal regions for the 3b (upper) and 4b (lower) event categories are shown separately
for the electron (left) and muon (right) channels.
The dotted lines $\wh^{20\GeV}$, $\wh^{60\GeV}$, 
illustrate the shapes of the signal template normalised to the SM cross section 
times a branching fraction $\mathcal{B}(\htobbbb) = 1$ and scaled by the factors indicated in the figure.
The horizontal error bars indicate the bin width. The hatched band in each bin represents the uncertainty in the total background estimation.
}
\label{fig:resultsbdt1}
\end{figure}       

\begin{figure}[t!]   
\centering   
\includegraphics[width=0.47\textwidth]{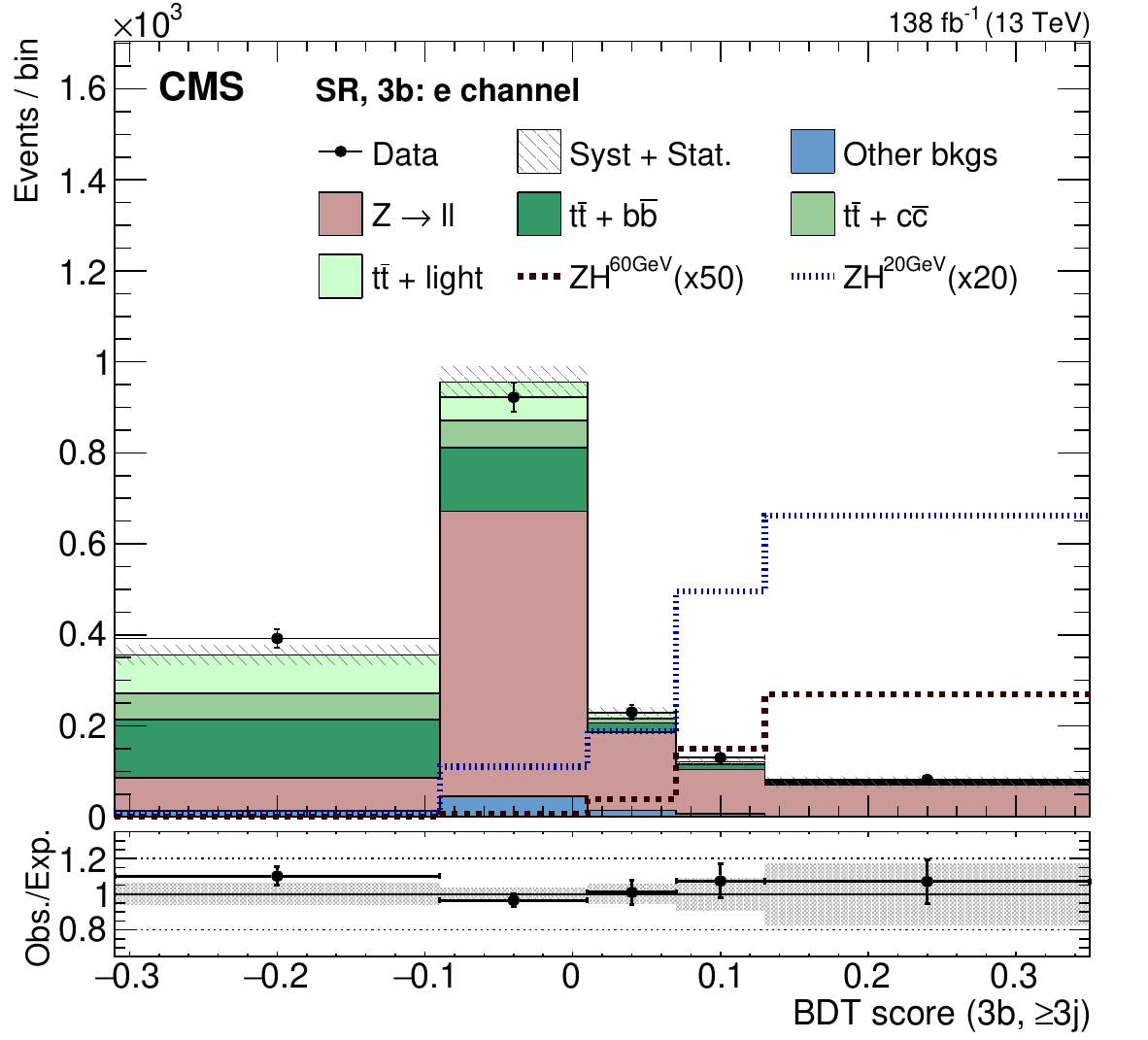} 
\hspace{3mm}    
\includegraphics[width=0.47\textwidth]{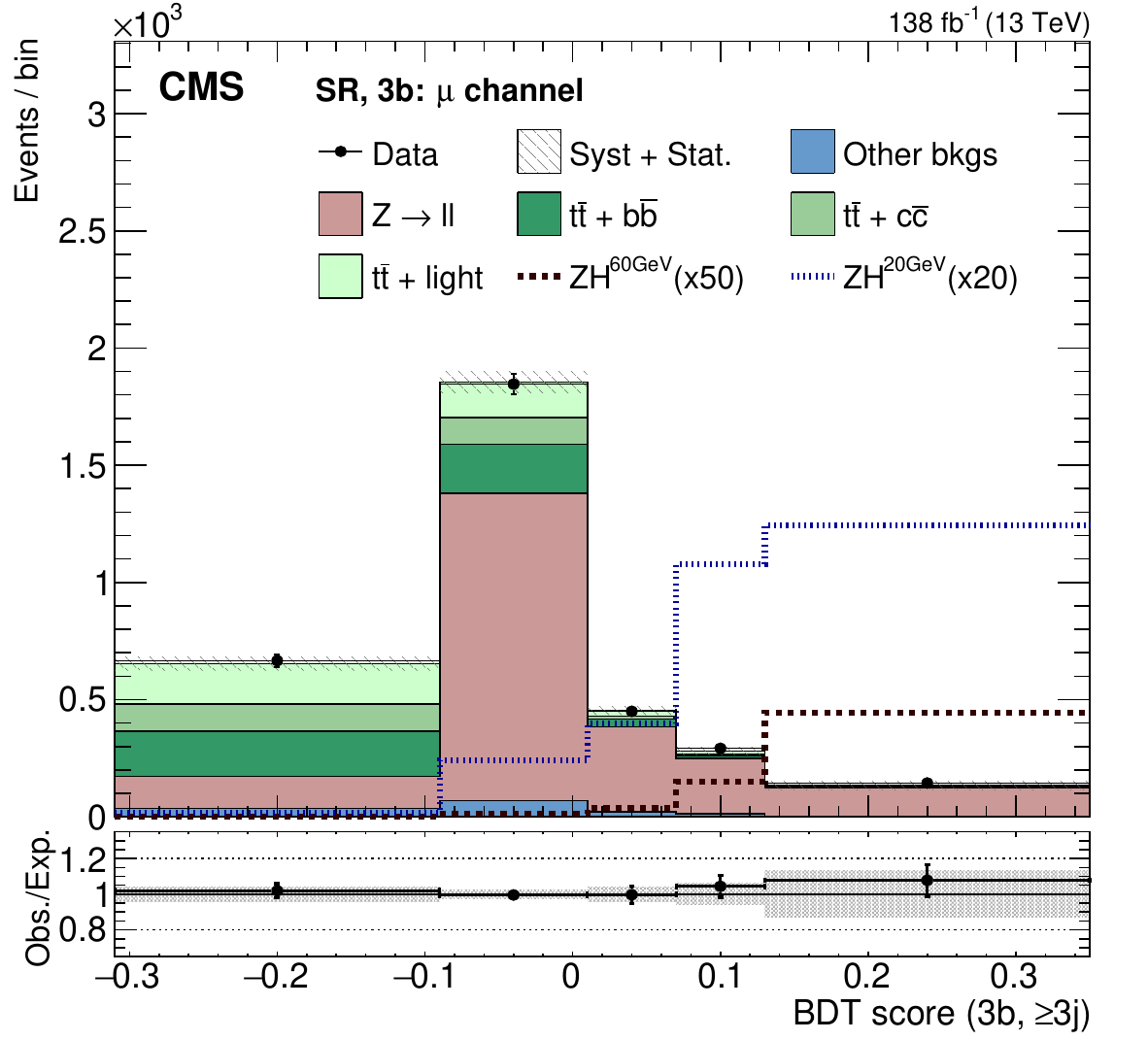}  
\includegraphics[width=0.47\textwidth]{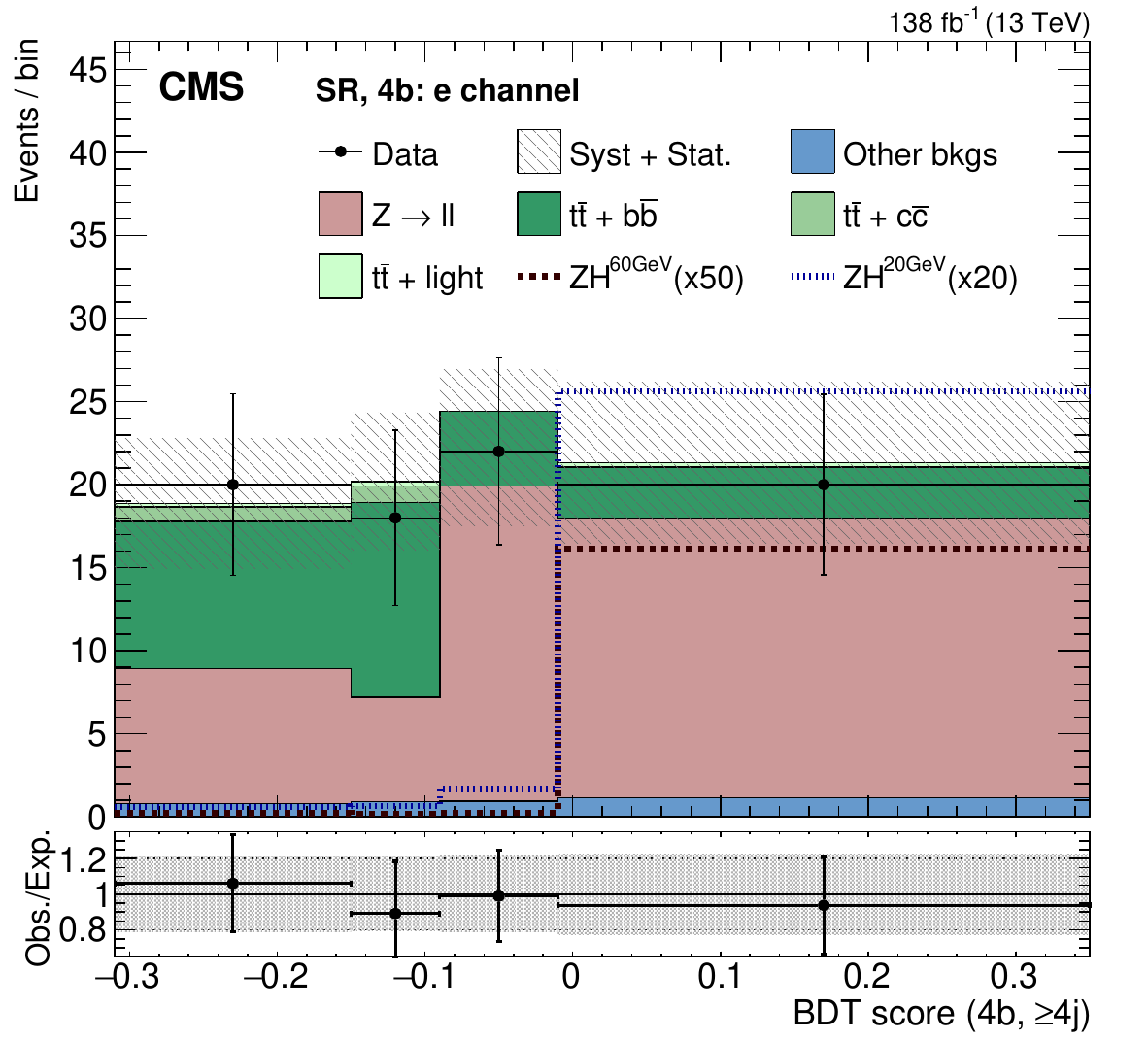} 
\hspace{3mm} 
\includegraphics[width=0.47\textwidth]{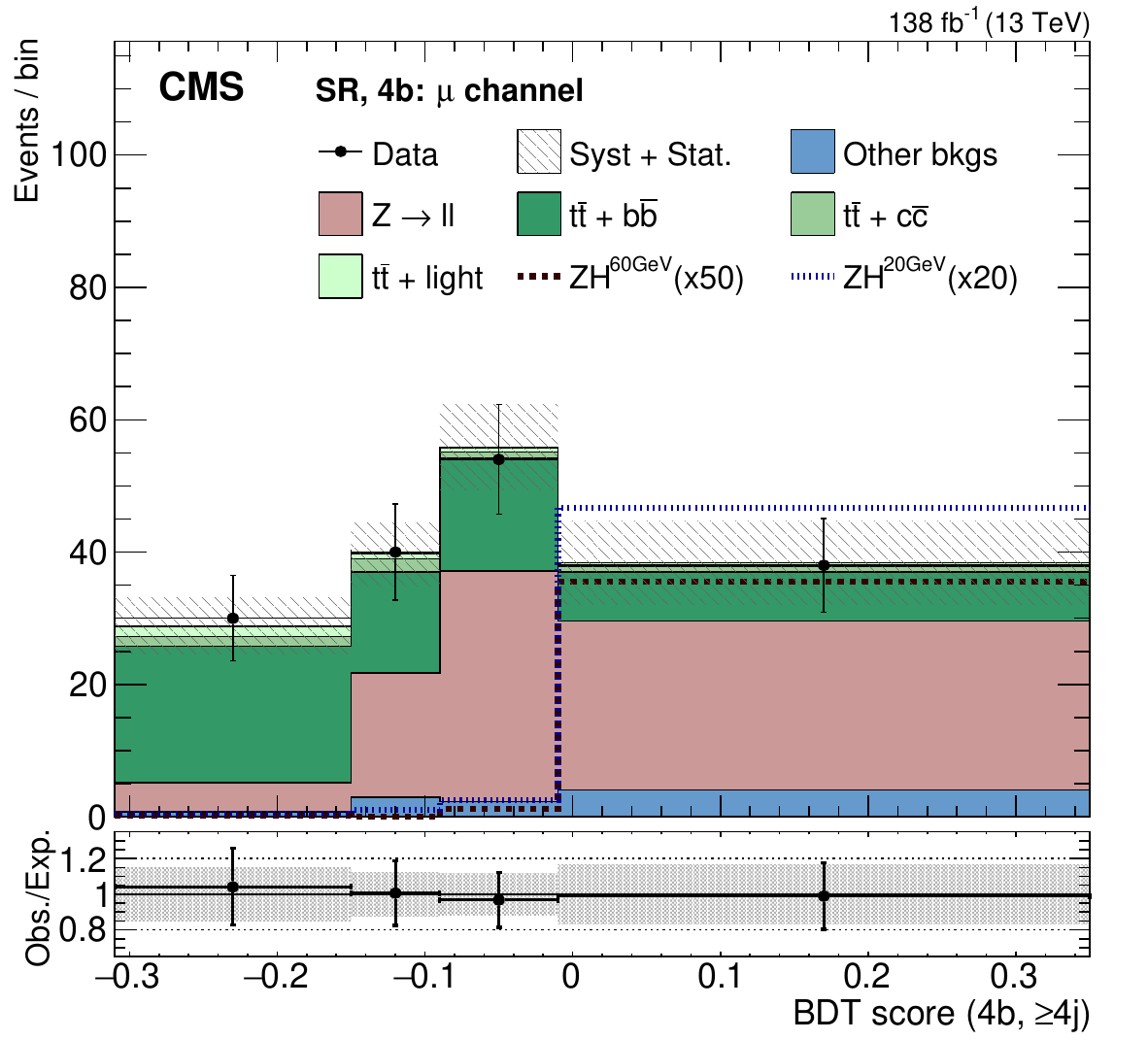} 
\vspace{4mm}  
\caption{
Post-fit BDT distributions in the \zh channel
extracted with the $\ma=60\GeV$ signal hypothesis.
Signal regions for the 3b (upper) and 4b (lower) event categories are shown separately
for the electron (left) and muon (right) channels.
The dotted lines $\zh^{20\GeV}$ and $\zh^{60\GeV}$, 
illustrate the shapes of the signal template normalised to the SM cross section 
times a branching fraction $\mathcal{B}(\htobbbb) = 1$ and scaled by the factors indicated in the figure.
The horizontal error bars indicate the bin width.The hatched band in each bin
represents the uncertainty in the total background estimation. 
}
\label{fig:resultsbdt2}
\end{figure}   

\begin{table}[t!]
\topcaption{
Signal-plus-background fit results for the 3b \wh and \zh signal regions
extracted with the $\ma=60\GeV$ signal hypothesis.
The lepton flavor (\Pe or \PGm) and BDT bin range (in square brackets)
are indicated in the column headings.
Signal yields corresponding to the expectation for SM $\PV\higgs$ production 
and $\mathcal{B}(\htobbbb) = 1$
 are shown for the $\ma=20$ and 60\GeV hypotheses.
The background uncertainties account for both systematic and statistical sources.
}
\centering
\cmsTable{
\begin{tabular}{ccccccc}
\hline
Process  & { \Pe  [1,3] }  & { \Pe  [4,4] }  &{ \Pe  [5,5] }  & { $\mu$ [1,3] }  & { $\mu$ [4,4] }  &{ $\mu$ [5,5] }  \\
\hline
\multicolumn{7}{c}{ $\PW\higgs$ signal region, 3b } \\
Other bkgs                 &      2288 $\pm$   97   &     32.8 $\pm$ 7.8  &        14.9 $\pm$    1.7   &      3570 $\pm$   160   &        50.4 $\pm$    3.6   &        17.2 $\pm$    2.4   \\
Multijet                   &      1370 $\pm$  160   &       72 $\pm$ 15   &          48 $\pm$     11   &      4400 $\pm$   380   &         128 $\pm$     18   &        43.4 $\pm$    8.0   \\ 
$\PZ \to \ell\ell$         &       476 $\pm$   46   &     17.7 $\pm$ 6.4  &         4.7 $\pm$    4.0   &       430 $\pm$    54   &          19 $\pm$     14   &         7.9 $\pm$    1.2   \\ 
$\PW \to \ell\nu$          &      2850 $\pm$  180   &      110 $\pm$ 37   &          49 $\pm  $   12   &      3670 $\pm$   150   &         250 $\pm$     38   &         121 $\pm$     23   \\ 
\ttbb                      &      5650 $\pm$  280   &     58.9 $\pm$ 5.4  &        20.5 $\pm$    2.4   &      6610 $\pm$   400   &        74.8 $\pm$    8.5   &        22.9 $\pm$    5.1   \\ 
\ttcc                      &   19\,080 $\pm$ 260    &      101 $\pm$ 12   &        38.4 $\pm$    4.1   &     30\,930 $\pm$ 410   &         162 $\pm$     35   &        47.0 $\pm$    6.2   \\ 
\ttlf                      &   11\,220 $\pm$ 190    &       98 $\pm$ 13   &        33.8 $\pm$    4.1   &     18\,010 $\pm$ 310   &       159.5 $\pm$    6.7   &        48.1 $\pm$    2.0   \\  [\cmsTabSkip]   \hline
Total bkg                  &   42\,940 $\pm$  200   &      490 $\pm$ 64   &         208 $\pm$     17   &     67\,630 $\pm$ 280   &         844 $\pm$     52   &         308 $\pm$     29   \\  [\cmsTabSkip] \hline   
Data                       &     42\,934            &       510           &         230                &     67\,688             &         874                &         326                \\ [\cmsTabSkip] \hline    
 Signal(20)      &        77.3 $\pm$    1.4   &        30.2 $\pm$   0.9   &        15.8 $\pm$    0.7   &       110.1 $\pm$ 1.6   &        46.2 $\pm$    1.1   &        23.5 $\pm$    0.8   \\ 
 Signal(60)      &       117.5 $\pm$    1.5   &        90.3 $\pm$   1.6   &        60.7 $\pm$    1.3   &       164.9 $\pm$ 1.7   &       125.5 $\pm$    1.9   &        83.0 $\pm$    1.5   \\  [\cmsTabSkip] \hline
\multicolumn{7}{c}{$\PZ\higgs$ signal region, 3b } \\
Other bkgs                 &        72 $\pm$   16   &   6.2 $\pm$   1.7   &         1.2 $\pm$    0.7   &         126 $\pm$  24   &        13.2 $\pm$    3.7   &         2.6 $\pm$    0.9   \\ 
$\PZ \to \ell\ell$         &       871 $\pm$   40   &  98.1 $\pm$   9.8   &          68 $\pm$     12   &        1811 $\pm$  58   &         237 $\pm$     16   &         123 $\pm$     16   \\ 
\ttbb                      &       288 $\pm$   49   &  10.0 $\pm$   2.4   &         2.9 $\pm$    1.1   &         434 $\pm$  57   &        11.8 $\pm$    2.1   &         3.4 $\pm$    5.5   \\ 
\ttcc                      &       128 $\pm$   33   &   2.6 $\pm$   2.9   &         2.2 $\pm$    4.9   &         242 $\pm$  51   &         5.2 $\pm$    1.7   &         1.9 $\pm$    5.3   \\ 
\ttlf                      &       180 $\pm$   11   &   4.2 $\pm$   0.3   &         2.3 $\pm$    0.6   &         346 $\pm$  16   &        12.5 $\pm$    0.8   &         2.4 $\pm$    0.6   \\  [\cmsTabSkip] \hline 

Total bkg                  &      1539 $\pm$   50   &  121 $\pm$     11   &            77 $\pm$   13   &       2959 $\pm$   65   &           280 $\pm$   17   &         134 $\pm$   18     \\ [\cmsTabSkip] \hline 
Data                       &      1544              &     130             &          82                &        2962             &         292                &         144                \\ [\cmsTabSkip] \hline 
 Signal(20)                &     6.2 $\pm$    0.6   &  9.9 $\pm$    0.4   &        13.3 $\pm$    0.4   &     13.1 $\pm$    0.8   &        21.6 $\pm$    0.5   &        24.9 $\pm$    0.6   \\ 
 Signal(60)                &     7.6 $\pm$    0.7   & 21.4 $\pm$    0.5   &        43.4 $\pm$    0.8   &     13.1 $\pm$    0.8   &        42.8 $\pm$    0.8   &        76.6 $\pm$    1.0   \\ 
\hline
\end{tabular}
}
\label{tab:yields-3b}
\end{table}

\begin{table}[t!]
\topcaption{
Signal-plus-background fit results for the 4b \wh and \zh signal regions
extracted with the $\ma=60\gev$ signal hypothesis.
The lepton flavor (\Pe or \PGm) and BDT bin range (in square brackets)
are indicated in the column headings.
Signal yields corresponding to the expectation for SM $\PV\higgs$ production 
and $\mathcal{B}(\htobbbb) = 1$
are shown for the $\ma=20$ and 60\GeV hypotheses.
The background uncertainties account for both systematic and statistical sources.
}
\centering
\begin{tabular}{ccccc}
\hline
Process  & { \Pe  [1,3] }  &{ \Pe  [4,4] }  & { $\mu$ [1,3] }  &{ $\mu$ [4,4] }  \\ 
\hline
\multicolumn{5}{c}{ $\PW\higgs$ signal region, 4b } \\ 
Other bkgs                 &        29.9 $\pm$    1.9   &         0.4 $\pm$    0.3   &        40.5 $\pm$    2.0   &         1.4 $\pm$    0.2   \\ 
Multijet                     &        23.2 $\pm$    3.8   &         3.7 $\pm$    1.3   &        74.5 $\pm$    9.2   &        12.1 $\pm$    3.0   \\ 
$\PZ \to \ell\ell$         &          21 $\pm$     11   &         2.5 $\pm$    4.7   &        12.4 $\pm$    1.9   &         3.2 $\pm$    3.1   \\ 
$\PW \to \ell\nu$          &        18.6 $\pm$    5.3   &           5 $\pm$     11   &          40 $\pm$     14   &        16.4 $\pm$    7.8   \\ 
\ttbb                      &         213 $\pm$     15   &         4.6 $\pm$    2.1   &         278 $\pm$     15   &         6.7 $\pm$    1.8   \\ 
\ttcc                      &       190.6 $\pm$    7.0   &         7.9 $\pm$    2.4   &         305 $\pm$     14   &         2.1 $\pm$    1.3   \\ 
\ttlf                      &        63.0 $\pm$    4.9   &         1.8 $\pm$    1.1   &        96.0 $\pm$    8.3   &         3.8 $\pm$    1.3   \\ [\cmsTabSkip] \hline 
Total bkg                  &         559 $\pm$     17   &          26 $\pm$     11   &         846 $\pm$     24   &        45.7 $\pm$    7.8   \\ [\cmsTabSkip]  \hline
Data                       &         560                &          30                &         838                &          48                \\ [\cmsTabSkip] \hline
 Signal(20)      &         1.0 $\pm$    0.5   &         1.9 $\pm$    0.2   &         1.4 $\pm$    0.6   &         4.2 $\pm$    0.7   \\ 
 Signal(60)      &         1.7 $\pm$    0.6   &        13.9 $\pm$    0.6   &         1.6 $\pm$    0.5   &        18.8 $\pm$    0.7   \\ [\cmsTabSkip] \hline
\multicolumn{5}{c}{ $\PZ\higgs$ signal region, 4b } \\ 
Other bkgs                 &         2.7 $\pm$    2.8   &         1.2 $\pm$    0.6   &         6.0 $\pm$    1.4   &         4.0 $\pm$    1.4   \\ 
$\PZ \to \ell\ell$         &        33.3 $\pm$    9.3   &        16.8 $\pm$    4.7   &        58.2 $\pm$    8.8   &        25.6 $\pm$    6.0   \\ 
\ttbb                      &        25.1 $\pm$    6.8   &         3.1 $\pm$    1.1   &        52.9 $\pm$    6.3   &         7.4 $\pm$    1.3   \\ 
\ttcc                      &      $-0.0$ $\pm$    3.1   &      $-0.4$ $\pm$    4.5   &         4.4 $\pm$    1.3   &         0.9 $\pm$    0.4   \\ 
\ttlf                      &         0.2 $\pm$    1.2   &         0.7 $\pm$    0.6   &         3.0 $\pm$    0.7   &         0.6 $\pm$    1.2   \\ [\cmsTabSkip] \hline  
Total bkg                  &        61.3 $\pm$    8.2   &        21.3 $\pm$    4.9   &         124 $\pm$     10   &        38.4 $\pm$    6.4   \\ [\cmsTabSkip] \hline  
Data                       &          60                &          20                &         124                &          38   \\ [\cmsTabSkip] \hline 
 Signal(20)      &         0.1 $\pm$    0.3   &         1.3 $\pm$    0.1   &         0.2 $\pm$    0.3   &         2.3 $\pm$    0.2   \\ 
 Signal(60)      &         0.2 $\pm$    0.3   &         6.6 $\pm$    0.3   &         0.3 $\pm$    0.3   &        12.0 $\pm$    0.4   \\ 
\hline
\end{tabular}

\label{tab:yields-4b}
\end{table}

No evidence is found for a significant excess of events above the SM background prediction. 
Model independent upper limits are placed, at 95\% confidence level (\CL),
on the branching fraction $\mathcal{B}(\htobbbb)$ as a function of $\ma$. 
The results for the signal are normalized to the NNLO SM cross sections of the Higgs boson production
$\sigma(\pp\to\PW\higgs)=1.37\unit{pb}$ and $\sigma(\pp\to\PZ\higgs)= 0.86\unit{pb}$.
The results are dominated by the statistical uncertainties.
The dominant systematic uncertainties are associated with the \Pb jet identification.
For $\ma = 60$\GeV,
the \Pb-tagging uncertainties on the yield vary between 6 and 9\% for the major background processes.
For the signal process, this uncertainty is around 8\%.
Other uncertainties lie below 5\%.

\begin{figure}[htp]   
\centering 
\includegraphics[width=0.65\textwidth]{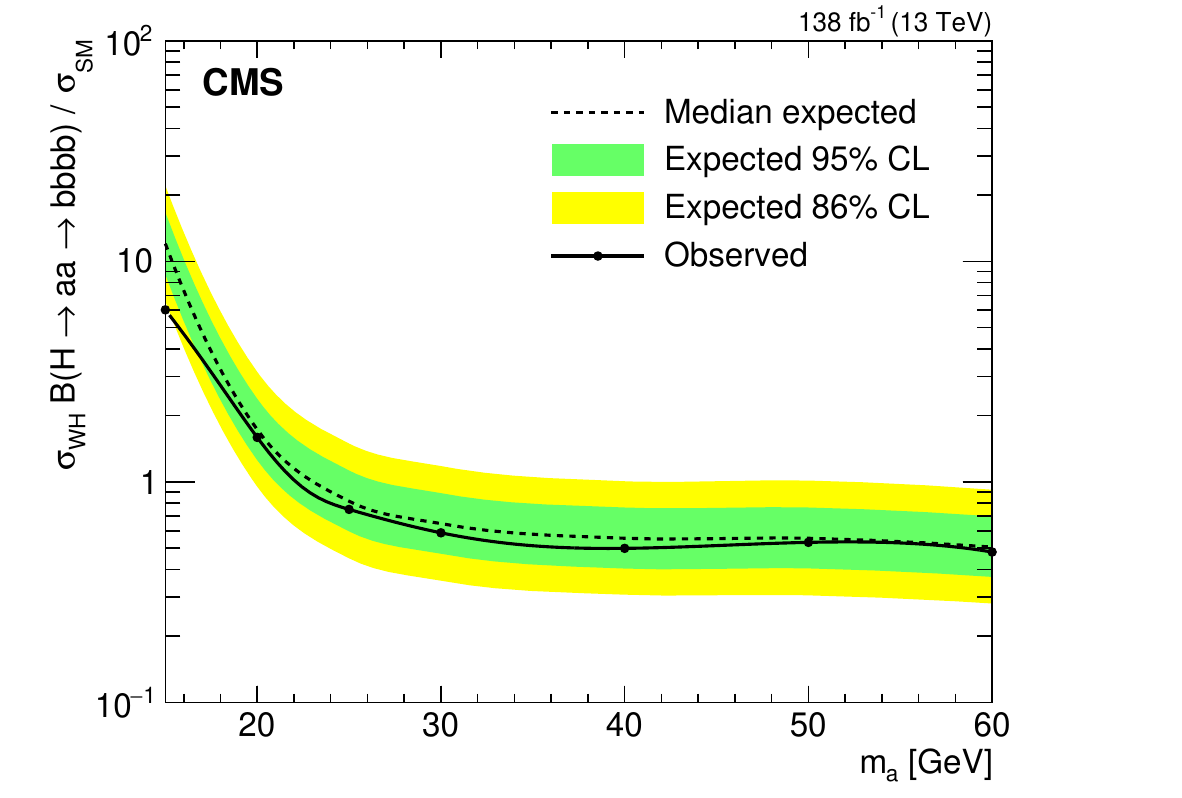}        
\includegraphics[width=0.65\textwidth]{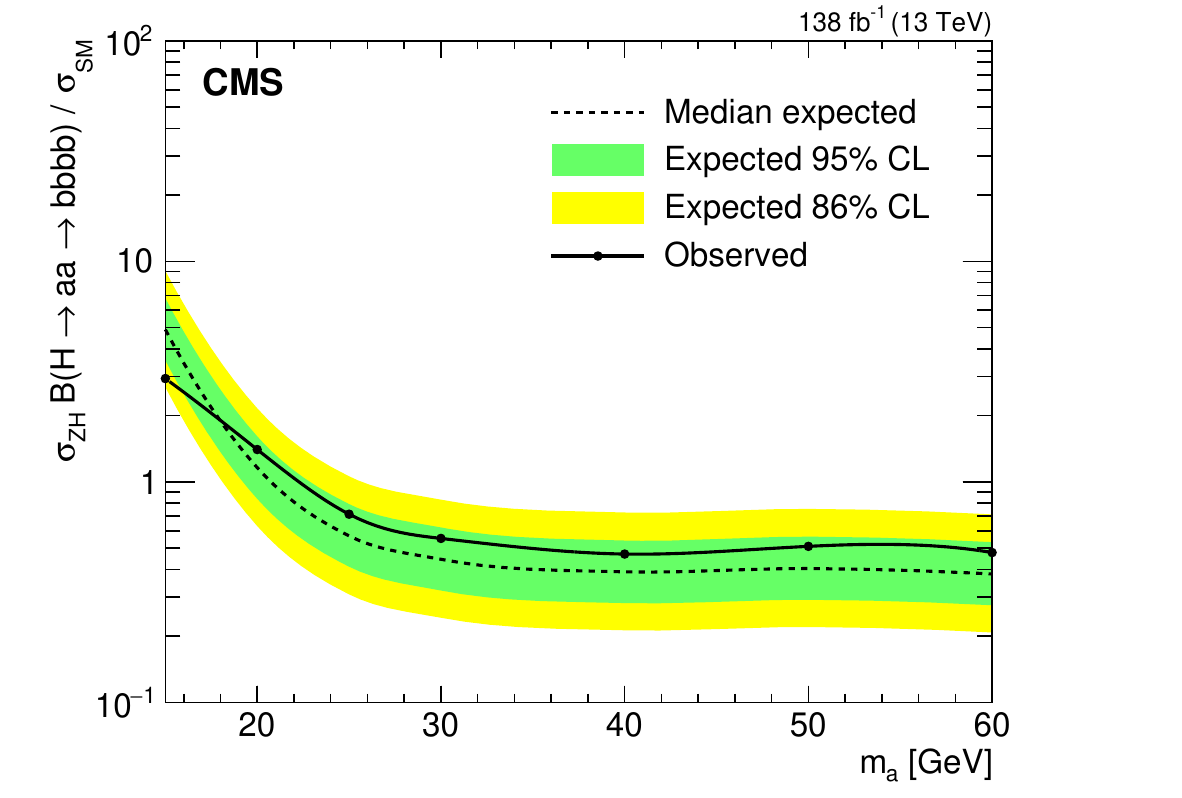} 
\includegraphics[width=0.65\textwidth]{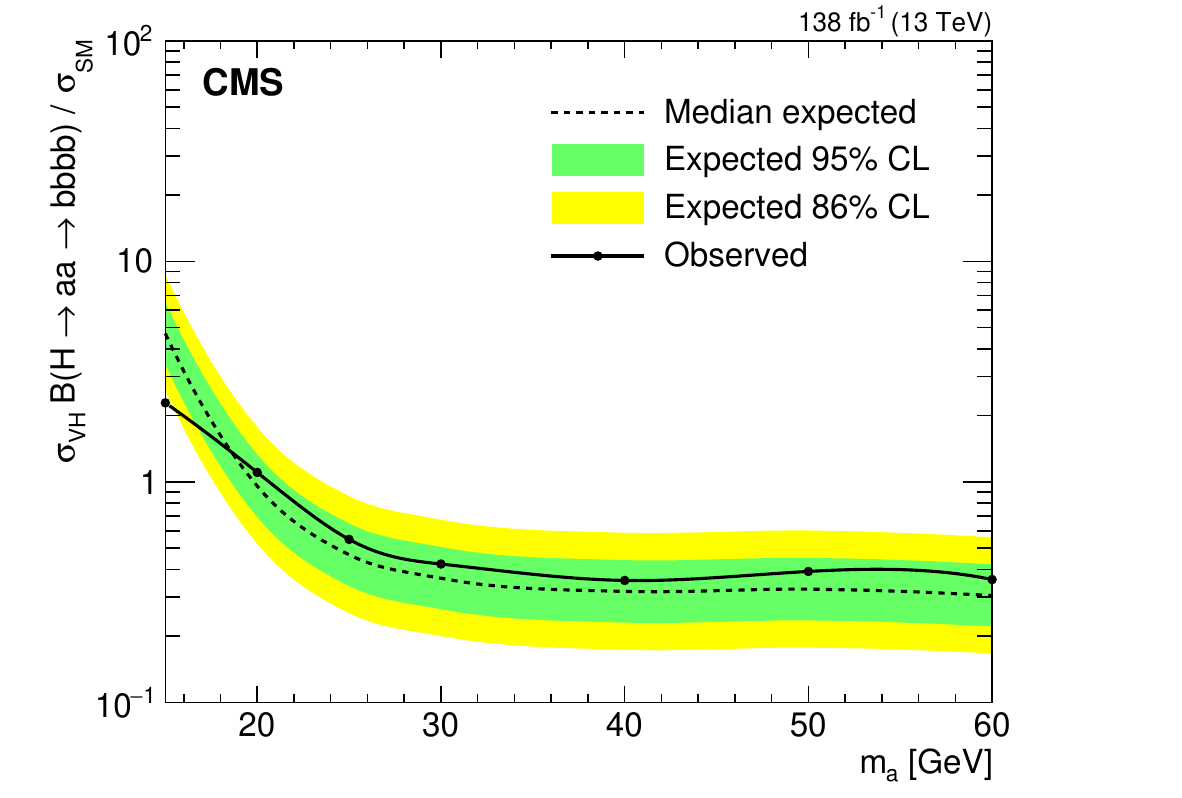}        
\caption{
Model independent 95\% CL upper limits on $\sigma_{\PV\higgs} \mathcal{B}(\htobbbb) / \sigma_{\mathrm{SM}}$ for
the \wh channel (upper), the \zh channel (middle), and the combination of both channels (lower),
where ``a'' is a new pseudoscalar particle decaying through $\Pa\to\bbbar$,
and  $\sigma_{\mathrm{SM}}$ is the SM Higgs boson production cross section.
}
\label{fig:limitsall}
\end{figure}

Figures~\ref{fig:resultsbdt1} and \ref{fig:resultsbdt2} show the post-fit shapes of the
BDT score variable
for the \wh and \zh channels, respectively,
where ``post-fit'' means that the parameter values determined in the likelihood fit are used.
The event yields for the various SM backgrounds
are presented in Tables~\ref{tab:yields-3b} and~\ref{tab:yields-4b}
for the 3b and 4b categories, respectively.
The results are presented in the bins of the BDT variable shown in
Figs.~\ref{fig:resultsbdt1} and \ref{fig:resultsbdt2}
(five bins for the 3b category and four bins for the 4b category).
For purposes of illustration,
signal yields corresponding to the expectation for SM $\PV\higgs$ production 
and $\mathcal{B}(\htobbbb) = 1$, for $\ma=20$ or 60\GeV are also shown.
The background uncertainties account for both systematic and statistical sources.
Figures~\ref{fig:limitsall} present model independent upper limits on the quantity
$\sigma_{\PV\higgs} \mathcal{B}(\htobbbb) / \sigma_{\mathrm{SM}}$ 
as a function of \ma, for the \wh and \zh channels separately
as well as their combination.
The limits are obtained
using the \CLs criterion and an asymptotic approximation
to the distribution of the profiled likelihood ratio test 
statistic~\cite{Read:2002hq,Cowan:2010js},
assuming SM predictions for the Higgs boson production cross section.
The limit curves are plotted by interpolating smoothly across the mass points.

In the \wh channel,
the observed upper limit on the $\mathcal{B}(\htobbbb)$ is found to be 0.48 (1.59) for $\ma= 60$ (20)\GeV,
compared to an expected limit of $0.51 \pm 0.16$ ($1.73 \pm 0.56$).
In the \zh channel,
the corresponding observed upper limit is 0.48 (1.40),
with an expected result of $0.38 \pm 0.13$ ($1.16 \pm 0.38$).
Our search is optimized for $\ma$ values between 20 and 60\GeV,
with signal sensitivity falling rapidly below 30\GeV. 
For $\ma\lesssim 30\GeV$,
the sensitivity is degraded in both the \wh and \zh channels 
because of the signal efficiency loss when reconstructing collimated \Pb jet pairs
from highly Lorentz-boosted pseudoscalar bosons. 

The observed limit on $\mathcal{B}(\htobbbb)$ for the combined \wh and \zh channels is $0.36$ ($1.10$)
for $\ma= 60$ (20)\GeV,
compared to an expected result of $0.31 \pm 0.10$ ($0.96 \pm 0.32$). 
It is possible to extract model dependent limits on $\mathcal{B}(\higgs\to\Pa\Pa)$ in the context of 2HDM+S models, 
predicting the branching fraction $\mathcal{B}(\Pa\to\bbbar)$, depending on the type and parameters of the model~\cite{Curtin:2013fra}.

\section{Summary} 
\label{sec:summary}

A search for exotic decays of the 125\GeV Higgs boson (\higgs)
to a pair of new light pseudoscalar bosons ({\Pa}),
followed by decay to four \Pb quark jets, \htobbbb,
is presented, using data recorded with the CMS detector.
The analysis is based on an integrated luminosity of \runtwolumi
collected at a center-of-mass energy of 13\TeV in 2016--2018.
The search is performed in the context of the associated \wh and \zh production
in which the \PW or \PZ boson decays leptonically,
$\PW\to\ell\nu$ or $\PZ\to\ell^+\ell^-$,
with $\ell$ an electron or muon.
No evidence for the targeted decay mode is observed.
The analysis obtains model independent upper limits at 95\% confidence level
on the branching fraction $\mathcal{B}(\htobbbb)$ of a SM-like Higgs boson.
The combined result for the associated \wh and \zh Higgs boson production excludes
branching fractions as low as 0.36 in the mass range $\ma$ between about 25\GeV and 60\GeV, assuming the SM
\wh and \zh cross-sections.
These results provide enhanced sensitivity,
in the mass range $\ma \gtrsim 20$\GeV, in complementary regions of the 2HDM+S model parameter space
compared to CMS searches in the 
$\PGm\PGm\PGt\PGt$ and $\Pb\Pb\ell\ell$ ($\ell=\PGm, \PGt$) final states. 

\begin{acknowledgments}
  We congratulate our colleagues in the CERN accelerator departments for the excellent performance of the LHC and thank the technical and administrative staffs at CERN and at other CMS institutes for their contributions to the success of the CMS effort. In addition, we gratefully acknowledge the computing centers and personnel of the Worldwide LHC Computing Grid and other centers for delivering so effectively the computing infrastructure essential to our analyses. Finally, we acknowledge the enduring support for the construction and operation of the LHC, the CMS detector, and the supporting computing infrastructure provided by the following funding agencies: SC (Armenia), BMBWF and FWF (Austria); FNRS and FWO (Belgium); CNPq, CAPES, FAPERJ, FAPERGS, and FAPESP (Brazil); MES and BNSF (Bulgaria); CERN; CAS, MoST, and NSFC (China); MINCIENCIAS (Colombia); MSES and CSF (Croatia); RIF (Cyprus); SENESCYT (Ecuador); ERC PRG, RVTT3 and MoER TK202 (Estonia); Academy of Finland, MEC, and HIP (Finland); CEA and CNRS/IN2P3 (France); SRNSF (Georgia); BMBF, DFG, and HGF (Germany); GSRI (Greece); NKFIH (Hungary); DAE and DST (India); IPM (Iran); SFI (Ireland); INFN (Italy); MSIP and NRF (Republic of Korea); MES (Latvia); LMTLT (Lithuania); MOE and UM (Malaysia); BUAP, CINVESTAV, CONACYT, LNS, SEP, and UASLP-FAI (Mexico); MOS (Montenegro); MBIE (New Zealand); PAEC (Pakistan); MES and NSC (Poland); FCT (Portugal); MESTD (Serbia); MCIN/AEI and PCTI (Spain); MOSTR (Sri Lanka); Swiss Funding Agencies (Switzerland); MST (Taipei); MHESI and NSTDA (Thailand); TUBITAK and TENMAK (Turkey); NASU (Ukraine); STFC (United Kingdom); DOE and NSF (USA).
    
  \hyphenation{Rachada-pisek} Individuals have received support from the Marie-Curie program and the European Research Council and Horizon 2020 Grant, contract Nos.\ 675440, 724704, 752730, 758316, 765710, 824093, 101115353, and COST Action CA16108 (European Union); the Leventis Foundation; the Alfred P.\ Sloan Foundation; the Alexander von Humboldt Foundation; the Science Committee, project no. 22rl-037 (Armenia); the Belgian Federal Science Policy Office; the Fonds pour la Formation \`a la Recherche dans l'Industrie et dans l'Agriculture (FRIA-Belgium); the Agentschap voor Innovatie door Wetenschap en Technologie (IWT-Belgium); the F.R.S.-FNRS and FWO (Belgium) under the ``Excellence of Science -- EOS" -- be.h project n.\ 30820817; the Beijing Municipal Science \& Technology Commission, No. Z191100007219010 and Fundamental Research Funds for the Central Universities (China); the Ministry of Education, Youth and Sports (MEYS) of the Czech Republic; the Shota Rustaveli National Science Foundation, grant FR-22-985 (Georgia); the Deutsche Forschungsgemeinschaft (DFG), under Germany's Excellence Strategy -- EXC 2121 ``Quantum Universe" -- 390833306, and under project number 400140256 - GRK2497; the Hellenic Foundation for Research and Innovation (HFRI), Project Number 2288 (Greece); the Hungarian Academy of Sciences, the New National Excellence Program - \'UNKP, the NKFIH research grants K 124845, K 124850, K 128713, K 128786, K 129058, K 131991, K 133046, K 138136, K 143460, K 143477, 2020-2.2.1-ED-2021-00181, and TKP2021-NKTA-64 (Hungary); the Council of Science and Industrial Research, India; ICSC -- National Research Center for High Performance Computing, Big Data and Quantum Computing, funded by the EU NexGeneration program (Italy); the Latvian Council of Science; the Ministry of Education and Science, project no. 2022/WK/14, and the National Science Center, contracts Opus 2021/41/B/ST2/01369 and 2021/43/B/ST2/01552 (Poland); the Funda\c{c}\~ao para a Ci\^encia e a Tecnologia, grant CEECIND/01334/2018 (Portugal); the National Priorities Research Program by Qatar National Research Fund; MCIN/AEI/10.13039/501100011033, ERDF ``a way of making Europe", and the Programa Estatal de Fomento de la Investigaci{\'o}n Cient{\'i}fica y T{\'e}cnica de Excelencia Mar\'{\i}a de Maeztu, grant MDM-2017-0765 and Programa Severo Ochoa del Principado de Asturias (Spain); the Chulalongkorn Academic into Its 2nd Century Project Advancement Project, and the National Science, Research and Innovation Fund via the Program Management Unit for Human Resources \& Institutional Development, Research and Innovation, grant B37G660013 (Thailand); the Kavli Foundation; the Nvidia Corporation; the SuperMicro Corporation; the Welch Foundation, contract C-1845; and the Weston Havens Foundation (USA).
\end{acknowledgments}

\bibliography{auto_generated}
\cleardoublepage \appendix\section{The CMS Collaboration \label{app:collab}}\begin{sloppypar}\hyphenpenalty=5000\widowpenalty=500\clubpenalty=5000
\cmsinstitute{Yerevan Physics Institute, Yerevan, Armenia}
{\tolerance=6000
A.~Hayrapetyan, A.~Tumasyan\cmsAuthorMark{1}\cmsorcid{0009-0000-0684-6742}
\par}
\cmsinstitute{Institut f\"{u}r Hochenergiephysik, Vienna, Austria}
{\tolerance=6000
W.~Adam\cmsorcid{0000-0001-9099-4341}, J.W.~Andrejkovic, T.~Bergauer\cmsorcid{0000-0002-5786-0293}, S.~Chatterjee\cmsorcid{0000-0003-2660-0349}, K.~Damanakis\cmsorcid{0000-0001-5389-2872}, M.~Dragicevic\cmsorcid{0000-0003-1967-6783}, P.S.~Hussain\cmsorcid{0000-0002-4825-5278}, M.~Jeitler\cmsAuthorMark{2}\cmsorcid{0000-0002-5141-9560}, N.~Krammer\cmsorcid{0000-0002-0548-0985}, A.~Li\cmsorcid{0000-0002-4547-116X}, D.~Liko\cmsorcid{0000-0002-3380-473X}, I.~Mikulec\cmsorcid{0000-0003-0385-2746}, J.~Schieck\cmsAuthorMark{2}\cmsorcid{0000-0002-1058-8093}, R.~Sch\"{o}fbeck\cmsorcid{0000-0002-2332-8784}, D.~Schwarz\cmsorcid{0000-0002-3821-7331}, M.~Sonawane\cmsorcid{0000-0003-0510-7010}, S.~Templ\cmsorcid{0000-0003-3137-5692}, W.~Waltenberger\cmsorcid{0000-0002-6215-7228}, C.-E.~Wulz\cmsAuthorMark{2}\cmsorcid{0000-0001-9226-5812}
% ...
\par}
\vskip\cmsinstskip
\dag:~Deceased\\
$^{1}$Also at Yerevan State University, Yerevan, Armenia\\
$^{2}$Also at TU Wien, Vienna, Austria\\
% ...
\end{sloppypar}
\end{document}